# Phase Field Model for Non-equilibrium Interface Conditions

Yue Li, Lei Wang∗, Junjie Li, Jincheng Wang, Zhijun Wang∗

State Key Laboratory of Solidification Processing, Northwestern Polytechnical University, Xi'an 710072, China

**Abstract**

This article presents a new phase-field formulation for non-equilibrium interface conditions in rapid phase transformations. With a particular way of defining concentration fields, the classical sharp and diffuse (thick) interface theories are unified into the present phase-field model. Each point inside the diffuse interface is comparable to the classical sharp interface model, governed by phase boundary migration and short-range atomic exchanges and connected by long-range diffusion. Their thermodynamic forces and fluxes are following the Onsager's reciprocal relation and consistent with the classical irreversible thermodynamics. Furthermore, we establish the balance of driving forces and energy dissipations for both the diffuse interface and the single point inside it. The model is then validated by rapid solidification of Al-Cu alloys. With an effective mobility of considering non-equilibrium long-range diffusion, the model represents the main characteristics of non-equilibrium interface kinetics, i.e., solute trapping and solute drag. Especially, we reproduce the complete trapping at the high-velocity regime and provide a thermodynamic-consistent description of the partial drag phenomena at the low-velocity region. Moreover, the dissipation analysis indicates that even the far-from-equilibrium diffuse interface is composed of numerous representative volume elements near the equilibrium, providing a new understanding of non-equilibrium interfaces.

∗ Corresponding author. lei.wang@nwpu.edu.cn (L. Wang)
∗ Corresponding author. zhjwang@nwpu.edu.cn (Z. Wang)

# I. Introduction

Rapid solidification has gained significant attention in the past few decades for its great potential in many industry technologies such as additive manufacturing [1–3], spray forming [4], welding [5], and synthesis of metastable compounds and glasses [6]. From a scientific aspect in material science and condensed matter, one critical topic related to rapid solidification is the thermodynamics of far-from-equilibrium interfaces during the mesoscale phase transformation, which is vital to quantitative understanding and prediction of the microstructural pattern formation under local non-equilibrium conditions, e.g., the rapidly solidified banded microstructures [7–10].

Solute trapping and solute drag are two main characteristics of far-from-equilibrium interfaces, which are widely supported by experiments [11–13] and atomic simulations [14–18] for rapid solidifications. The former means the composition deviation from the equilibrium phase diagram. The latter denotes the loss of thermodynamic driving force by trans-interface diffusion, which can significantly retard the interface motion at the low velocity regime. A quantitative model is required to reflect these two phenomena simultaneously.

The classical sharp interface theory has been extended into the non-equilibrium conditions [19–28] based on the classical irreversible thermodynamics [29–33]. In rapid solidification with negligible solid diffusion, they are usually called the *full drag* model to describe the solute drag effect of redistributing an amount of $c_l - c_s$ atoms back into the solid phase ($c_{i=s,l}$ is the concentration of solid or liquid). In addition, this is also comparable to Aziz et al.'s famous Continuous Growth Model (CGM) [34–37], in the case of linear thermodynamics. However, although consistent with the maximum entropy production principle (MEPP) [29–33], one serious problem of the *full-drag* model is that its predicted driving forces for different dissipative processes significantly differ from experiments [12,13,24] and atomic simulations [14–18]. Thus, an adjustable *partial-drag* coefficient is usually added to reproduce better the reality [24,38], which, however, is inconsistent with MEPP anymore [24]. Moreover, the extra phenomenological parameter will limit the predictability of other simulation techniques that are based on the sharp interface models as the standard (phase-field model [39]) or rule (cellular automation [40] and level-set [41]). Therefore, except for the atomic simulation, there is still a lack of a thermodynamically consistent description for interface kinetics far from the equilibrium.

The natural interfaces are usually diffusive, containing a transition zone with several atoms' widths [42–45]. Hence, modeling the diffuse interface may provide new insight and more details about the non-equilibrium interfaces. Phase-field theory (PF) has been extensively developed in the past two decades [46–51] as a representative diffuse interface method. If the interface width is treated as a physical entity consistent with the natural interfaces, the classical continuum PF models should inherently reflect the intrinsic properties of diffuse interfaces [49,50]. However, less scientific attention focuses on the internal dissipative processes and their thermodynamics. For instance, the diffuse interface in the classical multi-phase-field (MPF) models [52] is assumed as a mixture of two phases with their individual concentration fields. However, the assumption of pointwise equality of diffusion potential [53] imposes that different phases contacting in the diffuse interface will unrealistically exchange atoms so fast to reach a near (local)-equilibrium condition.

To overcome this limitation, Steinbach and Zhang et al. [54,55] have proposed a novel finite dissipation model by introducing a kinetic equation for the atomic exchange. The predicted solute trapping [$k(V)$] agrees well with experiments, but the relationship between interface velocity and thermodynamic driving force, i.e., the atomic simulated partial drag phenomenon [14–18], has not been reported. Moreover, because this model is not analytically tractable, it is unclear how this model compares to the solute drag force of classical diffuse (thick) interface theories by Cahn [56], Hillert and Sundman [57]. In contrast, Wang et al. [58,59] have proposed another treatment for representing short-range atomic redistribution based on the Maximal Entropy Production Principle (MEPP) [29–33]. Wang et al.'s model elegantly follows Onsager's reciprocal relationship [29,30] and reproduces Hillert and Sundman's solute drag of diffuse interface [57], yet its concentration fields usually diverge at the phase boundaries [58]. Therefore, a satisfactory PF (diffuse interface) model independent from the sharp interface model is still lacking.

To address these issues, we present a new non-equilibrium PF model by separating concentration fields in a new way in the classical MPF framework. Specifically, each concentration field is further divided into conserved and non-conserved, which can kinetically couple the short-range atomic exchange and long-range diffusion through a diffusion equation with additional source or sink. Then, the diffuse interface is viewed as an integral of numerous sharp interfaces. A "two-part" mechanism for each sharp interface is proposed to balance the driving forces and energy dissipation for displacive and diffusional processes. Moreover, integrating the whole interface can recover the solute drag of classical diffuse interface theory [56,57], and reproduce the atomic simulated partial drag phenomenon.

The remainder of this article is organized as follows: In Sec. II, the present non-equilibrium diffuse interface model is established based on the classical PF framework and extremum principles of irreversible thermodynamics. In Sec. III, we mainly focus on the thermodynamics of non-equilibrium diffuse interface, especially the balance of driving forces and energy dissipation for different dissipative processes. In Sec. V, the present PF model is applied in the rapidly solidified Al-Cu alloy to reproduce the atomic simulated solute trapping and partial solute drag effects. The main conclusions are summarized in Sec. VI. In the Appendix, we reinterpret the previous sharp interface models in a new way for better comparison with the present diffuse interface theory.

## II. The Phase Field Model

### A. Description of diffusion interface

The present phase-field model is based on the following understanding of dissipative processes within the diffuse interface. As shown in Figure. 1(a), we consider the diffuse interface as a continuum collection of numerous representative volume elements (RVEs). In each RVE, there are $\alpha$ and $\gamma$ separated by a distinct phase boundary, which represents the abrupt transition of the lattice structure. Thus, this boundary is called the "sharp interface" in this work. Note that the actual interface is a uniform mixture of RVEs, and Figure. 1(a) is just an averaged version, representing the gradual transition of phase fractions at the different positions.

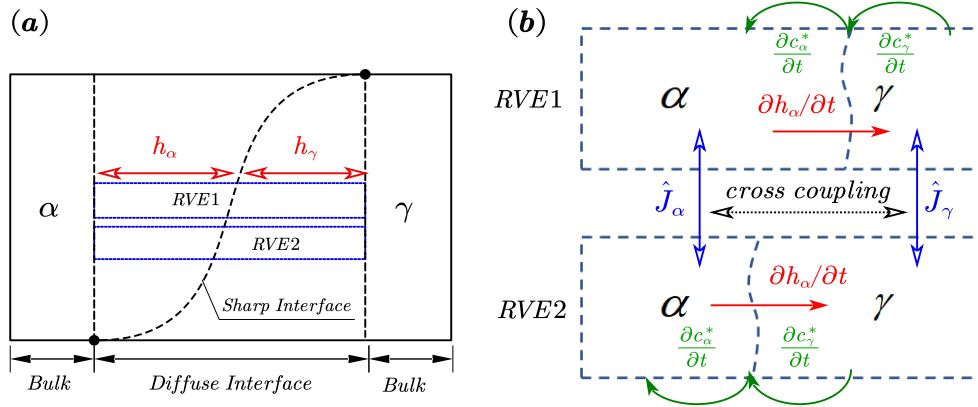

FIG. 1. (a) Diffuse interface and the internal representative volume elements (RVEs); (b) Three dissipative processes between and within RVEs, where the blue arrows means the long-range diffusion in the same phase while the green arrows are the short-range exchange between different phases.

During the $\alpha-\gamma$ phase transformation, the migration of the diffuse interface [Figure. 1(a)] is accomplished by those of sharp interfaces in RVEs [red arrows in Figure. 1(b)]. At the same time, there

5 / 52

are short-range atomic jumps between different phases to adjust the composition, i.e., the green arrows in Figure. 1(b). This scenario is similar to the classical sharp interface model shown in Figure. A1 of the Appendix, thus, the short-range atomic redistribution is referred to as "trans-sharp-interface diffusion" currently. In addition, due to the spatial variation of diffusion potential through the diffuse interface, there is a long-range diffusion in the same phase across different RVEs, defined as "trans-diffuse-interface diffusion", as shown by blue arrows in Figure. 1(b). It should note that the driving forces for these two types of diffusion are different. Analogous to Aziz's model based on chemical rate theory [34–37], the trans-sharp-interface diffusion, atoms transferring from $\alpha(\gamma)$ lattice to $\gamma(\alpha)$ lattice, is driven by diffusion potential itself, which is also similar to Eq. (A1) in the classical sharp interface models [24–27]. In contrast, the trans-diffuse-interface diffusion is driven by the gradient of diffusion potential.

There are two examples supporting the former understanding of diffuse interface. The first example is the step flow growth of crystal surfaces exposed to the gas phase [60], where the diffuse interface is composed of many discontinuous terraces. The adatoms' adsorption and desorption between terraces and the gas are analogous to the present trans-sharp-interface diffusion. Simultaneously, long-range diffusions exist within crystal surface and the gas. The other is a recent neutron total scattering characterization of the interface structure in Mg-Al alloy [61]. The experiment reveals that the interface consists of short-range order cells (or groups) of two phases, similar to the present RVEs. The overall interface migration is accomplished by atomic-level displacive transformation of these cells, aided by diffusion between cells. The similarity to our model in phase transformation mechanisms will be further discussed in Sec. III.

**B. Derivation of phase-field model**

The model derivations start from a classical free energy functional of a binary two-phase system, i.e., [53]

$$F = \int_\Omega \left\{ \sum_{i=\alpha,\gamma} h_i(\phi)\left(f_i(c_i)/v_m\right) + wg(\phi) + \frac{\kappa^2}{2}(\nabla\phi)^2 \right\} d\Omega, \quad (1)$$

in which $h_\alpha(\phi) = \phi^2(3-2\phi)$ is used to interpolate the bulk free energy densities of two phases $f_i(c_i)/v_m$, and the molar volume ($v_m$) for two phases are assumed to be the same for simplicity. $c_\alpha$ and $c_\gamma$ are the separated concentration fields. $g(\phi) = \phi^2(1-\phi)^2$ is the double well potential with minima at $\phi = 0, 1$, and $w$ is the barrier height. $\kappa^2$ is the gradient energy coefficient. The overall concentration is $c = h_\alpha c_\alpha + h_\gamma c_\gamma$, whose constraint to the systematic evolutions can be considered through an additional functional

$$F_{ad} = \lambda_1 \int_\Omega \left\{ c - \left( h_\alpha c_\alpha + h_\gamma c_\gamma \right) \right\} d\Omega, \quad (2)$$

where $\lambda_1$ is the Lagrange multiplier.

Based on the former analysis of diffuse interface in Sec. II. A, it is natural to know the separate concentration fields $c_\alpha$ and $c_\gamma$ should be non-conserved, instead of Wang et al.'s conserved assumptions [58]. The trans-sharp-interface diffusion plays a role of source or sink for the trans-diffuse-interface diffusion. Therefore, we further divide $c_{i=\alpha,\gamma}$ into a conserved $\hat{c}_i$ and a non-conserved $c_i^*$,

$$c_{i=\alpha,\gamma} = \hat{c}_i + c_i^*, \quad (3)$$

whose evolution can be represented by the diffusion equation with additional sources or sinks,

$$\frac{\partial c_{i=\alpha,\gamma}}{\partial t} = -v_m \nabla \cdot \hat{J}_i + \frac{\partial c_i^*}{\partial t}, \quad (4)$$

where $\hat{J}_i$ and $\partial c_i^*/\partial t$ ($i = \alpha, \gamma$) are the fluxes of trans-diffuse-interface diffusion and trans-sharp-interface diffusion introduced in the Sec. II. A, as marked in Figure. 1(b). In contrast, the temporal evolution of overall concentration is still conserved, i.e.,

$$\frac{\partial c}{\partial t} = -v_m \nabla \cdot J. \tag{5}$$

Substituting Eqs. (4), (5) into Eqs. (1), (2) and applying the divergence theorem, we have

$$\dot{F} = \int_\Omega \left\{ \frac{\partial \phi}{\partial t} \frac{\delta F}{\delta \phi} + \sum_{i=\alpha,\gamma} \left( \hat{J}_i \cdot v_m \nabla \frac{\delta F}{\delta c_i} \right) + \sum_{i=\alpha,\gamma} \frac{\partial c_i^*}{\partial t} \frac{\delta F}{\delta c_i} \right\} d\Omega, \tag{6}$$

and

$$\dot{F}_{ad} = \lambda_1 \int_\Omega \left\{ v_m \nabla h_\alpha \left( \hat{J}_\alpha - \hat{J}_\gamma \right) + h_\alpha \frac{\partial c_\alpha^*}{\partial t} + h_\gamma \frac{\partial c_\gamma^*}{\partial t} + \frac{\partial h_\alpha}{\partial \phi} \frac{\partial \phi}{\partial t} (c_\alpha - c_\gamma) \right\} d\Omega, \tag{7}$$

in which the surface integrals are neglected for a closed system. Similar to the classical sharp interface models in Appendix, note that it is the trans-sharp-interface diffusion, together with phase boundary migration, that determines mass conservation after phase transformation in RVEs. Hence, there should be $\hat{J}_\alpha = \hat{J}_\gamma = \hat{J}$ transforming Eqs. (6) and (7) into

$$\dot{F} = \int_\Omega \left\{ \frac{\partial \phi}{\partial t} \frac{\delta F}{\delta \phi} + \hat{J} \cdot \sum_{i=\alpha,\gamma} v_m \nabla \frac{\delta F}{\delta c_i} + \sum_{i=\alpha,\gamma} \left( \frac{\partial c_i^*}{\partial t} \frac{\delta F}{\delta c_i} \right) \right\} d\Omega, \tag{8}$$

and

$$\dot{F}_{ad} = \lambda_1 \int_\Omega \left\{ h_\alpha \frac{\partial c_\alpha^*}{\partial t} + h_\gamma \frac{\partial c_\gamma^*}{\partial t} + \frac{\partial h_\alpha}{\partial \phi} \frac{\partial \phi}{\partial t} (c_\alpha - c_\gamma) \right\} d\Omega, \tag{9}$$

where Eq. (9) relates three non-conserved fluxes as Eq. (A4) of the classical sharp interface model in Appendix. Moreover, the net flux $J$ follows the mixture rule $J = h_\alpha \hat{J}_\alpha + h_\gamma \hat{J}_\gamma = \hat{J}$ and thus we have $c = \hat{c}_\alpha = \hat{c}_\gamma$.

To solve the governing equations constrained by Eq. (9), the maximal entropy production principle (MEPP) [31] is employed, which is equivalent to Onsager's original variation principle [29,30] for linear irreversible thermodynamics. For an isothermal-isobaric-isotropic system, MEPP for linear cases was further simplified by Svoboda et al. [32,33] with a simplified quadratic form of energy dissipation, usually called the thermodynamic extremal principle (TEP). Nowadays, MEPP or TEP has been successfully applied in modeling a variety of microstructure evolutions in materials area, e.g., diffusion [62], precipitation [63], diffusional phase transformation [26,27], creep [64], grain growth and coarsening [65,66]. As for the phase-field modeling, the classical Allen-Cahn [67] and Cahn-Hilliard [68] equations can also be recovered [69,70] if there is no additional constraint.

Following MEPP or TEP in the linear thermodynamics [32,33], the total free energy dissipation for four fluxes in Eq. (8) can be given by (Note that $\hat{J}_\alpha = \hat{J}_\gamma = \hat{J} = J$)

$$Q = \int_\Omega \left\{ \frac{1}{M_\phi}\left(\frac{\partial \phi}{\partial t}\right)^2 + \sum_{i=\alpha,\gamma} \frac{J^2}{M_i(\phi)} + \sum_{i=\alpha,\gamma} \frac{1}{M_i^*(\phi)}\left(\frac{\partial c_i^*}{\partial t}\right)^2 \right\} d\Omega, \tag{10}$$

in which $M_\phi$, $M_i(\phi)$, $M_i^*(\phi)$ are the mobilities of phase-field migration, trans-diffuse-interface diffusion, and trans-sharp-interface diffusion, respectively. For simplicity of later derivations, the inverse interpolation of $M_i(\phi)$ can be represented by

$$\frac{1}{M_c(\phi)} = \frac{1}{M_\alpha(\phi)} + \frac{1}{M_\gamma(\phi)}. \tag{11}$$

In order to ensure the maximal entropy production [32,33], the systematic evolutions should follow the relation $\dot{F} + \dot{F}_{ad} + Q = 0$, thus, the thermodynamic variations are given by

$$\delta\left[Q + \lambda_2\left(\dot{F} + \dot{F}_{ad} + Q\right)\right]\bigg|_{\frac{\partial \phi}{\partial t}, J, \frac{\partial c_i^*}{\partial t}} = 0, \tag{12}$$

where the Lagrange multiplier $\lambda_2 = -2$ is solved firstly [32,33], and then we have

$$\frac{1}{M_\phi}\frac{\partial \phi}{\partial t} + \frac{\delta F}{\delta \phi} + \lambda_1 \frac{\partial h_\alpha}{\partial \phi}(c_\alpha - c_\gamma) = 0, \tag{13}$$

$$\frac{1}{M_\alpha^*(\phi)}\frac{\partial c_\alpha^*}{\partial t} + \frac{\delta F}{\delta c_\alpha} + \lambda_1 h_\alpha \frac{\partial c_\alpha^*}{\partial t} = 0, \tag{14}$$

$$\frac{1}{M_\gamma^*(\phi)}\frac{\partial c_\gamma^*}{\partial t}+\frac{\delta F}{\delta c_\gamma}+\lambda_1 h_\gamma \frac{\partial c_\gamma^*}{\partial t}=0, \tag{15}$$

$$J=-M_c(\phi)v_m \nabla\left(\frac{\delta F}{\delta c_\alpha}+\frac{\delta F}{\delta c_\gamma}\right). \tag{16}$$

The trans-diffuse-interface diffusion $J$ is independent of the other three non-conserved fluxes participating in the mass conservation of phase transformation within RVEs, which, according to Eq. (9) is given by

$$h_\alpha \frac{\partial c_\alpha^*}{\partial t}+h_\gamma \frac{\partial c_\gamma^*}{\partial t}=-\frac{\partial h_\alpha}{\partial \phi}\frac{\partial \phi}{\partial t}(c_\alpha-c_\gamma). \tag{17}$$

Combining Eqs. (13)-(15) and Eq. (17), the first Lagrange multiplier $\lambda_1$ is then solved as

$$\lambda_1=-\frac{1}{\bar{M}^*}\left[\frac{\partial h_\alpha}{\partial \phi}\frac{\partial \phi}{\partial t}(c_\gamma-c_\alpha)+h_\alpha M_\alpha^*(\phi)\frac{\delta F}{\delta c_\alpha}+h_\gamma M_\gamma^*(\phi)\frac{\delta F}{\delta c_\gamma}\right], \tag{18}$$

where

$$\bar{M}^*=h_\alpha^2 M_\alpha^*(\phi)+h_\gamma^2 M_\gamma^*(\phi). \tag{19}$$

Submitting Eqs. (18) and (19) into Eqs. (13)-(15), the governing equations for three non-conserved fluxes yield

$$\frac{1}{M_\phi^{eff}}\frac{\partial \phi}{\partial t}=-\frac{\delta F}{\delta \phi}+\frac{\partial h_\alpha}{\partial \phi}\frac{(c_\gamma-c_\alpha)}{\bar{M}^*}\left[h_\alpha^2 M_\alpha^*(\phi)\tilde{\mu}_\alpha+h_\gamma^2 M_\gamma^*(\phi)\tilde{\mu}_\gamma\right] \tag{20}$$

where

$$\frac{1}{M_\phi^{eff}}=\frac{1}{M_\phi}+\left(\frac{\partial h_\alpha}{\partial \phi}\right)^2\frac{(c_\gamma-c_\alpha)^2}{\bar{M}^*}, \tag{21}$$

and

$$\frac{\partial c_\alpha^*}{\partial t}=\frac{h_\gamma}{v_m}\frac{h_\alpha h_\gamma M_\alpha^*(\phi)M_\gamma^*(\phi)}{\bar{M}^*}(\tilde{\mu}_\gamma-\tilde{\mu}_\alpha)+\frac{M_\alpha^*(\phi)h_\alpha}{\bar{M}^*}\frac{\partial h_\alpha}{\partial \phi}\frac{\partial \phi}{\partial t}(c_\gamma-c_\alpha), \tag{22}$$

$$\frac{\partial c_\gamma^*}{\partial t}=\frac{h_\alpha}{v_m}\frac{h_\alpha h_\gamma M_\alpha^*(\phi)M_\gamma^*(\phi)}{\bar{M}^*}(\tilde{\mu}_\alpha-\tilde{\mu}_\gamma)+\frac{M_\gamma^*(\phi)h_\gamma}{\bar{M}^*}\frac{\partial h_\alpha}{\partial \phi}\frac{\partial \phi}{\partial t}(c_\gamma-c_\alpha), \tag{23}$$

in which $\tilde{\mu}_{i=\alpha,\gamma} = (\delta F/\delta c_i)/h_i$ is the diffusion potential.

## C. Atomic mobility and model simplification

In Sec. II. B, Eqs. (16), (22), and (23) give the governing equations of the trans-diffuse-interface and trans-sharp-interface diffusions. However, the detailed expression of their atomic mobilities are still unknown, especially the short-range $M_i^*(\phi)$.

As employed in the Wheeler-Boettinger-McFadden (WBM) model [71] and Kim-Kim-Suzuki (KKS) model [53], the atomic mobility for long-range diffusion $M_c(\phi)$ is usually assumed to follow a linear interpolation

$$M_c(\phi) = h_\alpha M_\alpha + (1-h_\alpha)M_\gamma, \tag{24}$$

where $M_{i=\alpha,\gamma} = (D_i/v_m)(\partial \tilde{\mu}_i/\partial c_i)^{-1}$ is the atomic mobility for diffusion in bulks, $D_i$ is the diffusion coefficient. One way of recovering Eq. (24) from Eq. (11) is to assume the atomic mobility of trans-diffuse-interface diffusion, i.e., $M_\alpha(\phi)$ and $M_\gamma(\phi)$, follow the following relations

$$M_\alpha(\phi) = \frac{M_\alpha}{P(\phi)}; \quad M_\gamma(\phi) = \frac{M_\gamma}{1-P(\phi)}, \tag{25}$$

where the interpolation function $P(\phi)$ is expressed as [72]

$$P(\phi) = \frac{h_\alpha M_\alpha}{h_\alpha M_\alpha + (1-h_\alpha)M_\gamma}. \tag{26}$$

The atomic mobilities for trans-sharp-interface diffusion $M_\alpha^*(\phi)$ and $M_\gamma^*(\phi)$, can now be obtained from the comparison with the classical sharp interface model in Appendix. Using the standard coordinate transformation at the steady-state, i.e., $\partial \phi/\partial t \to -V_n(\partial \phi/\partial x)$, where $V_n$ is the interface velocity, the mass conservation Eq. (17) can be rewritten as

$$\frac{h_\alpha}{(\partial h_\alpha/\partial x)v_m}\frac{\partial c_\alpha^*}{\partial t} - \frac{h_\gamma}{(\partial h_\gamma/\partial x)v_m}\frac{\partial c_\gamma^*}{\partial t} = \frac{V_n}{v_m}(c_\alpha - c_\gamma), \tag{27}$$

which is consistent with Eq. (A3) of the classical sharp interface model in Appendix. Then, comparing Eqs. (22), (23) and Eqs. (A7), (A8), it is known that $M_\alpha^*(\phi)$ and $M_\gamma^*(\phi)$ should follow

$$M_\alpha^*(\phi) = \frac{M_\alpha(\phi) v_m^2}{a h_\alpha^2}\left(-\frac{\partial h_\alpha}{\partial x}\right) q_\alpha(\phi), \tag{28}$$

$$M_\gamma^*(\phi) = \frac{M_\gamma(\phi) v_m^2}{a h_\gamma^2}\left(\frac{\partial h_\gamma}{\partial x}\right) q_\gamma(\phi). \tag{29}$$

$q_\alpha(\phi)$ and $q_\gamma(\phi)$ are possible interpolation functions not changing the dimensions. In general, the gradient of $\phi$ deviates slightly from its equilibrium state [53], thus, Eqs. (28) and (29) can further denoted by

$$M_\alpha^*(\phi) = \frac{M_\alpha(\phi) v_m^2}{a} \frac{\sqrt{2w}}{\kappa}\left[\frac{\partial h_\alpha}{\partial \phi} \frac{q_\alpha(\phi)\phi(1-\phi)}{h_\alpha^2}\right], \tag{30}$$

$$M_\gamma^*(\phi) = \frac{M_\gamma(\phi) v_m^2}{a} \frac{\sqrt{2w}}{\kappa}\left[\frac{\partial h_\alpha}{\partial \phi} \frac{q_\gamma(\phi)\phi(1-\phi)}{h_\gamma^2}\right], \tag{31}$$

where $a$ is the interatomic spacing, $w$ and $\kappa$ are the coefficients in Eq. (1). It is evident the non-dimensional [ ] terms are two additional degrees of freedom in this model. To eliminate one degree of freedom and simplify the governing equations, it is convenient to let [ ] $= Cons$. The numerical application in later sections indicates that $Cons = 1$ is a good choice, i.e.,

$$M_\alpha^*(\phi) h_\alpha = M_\gamma^*(\phi) h_\gamma = M_c(\phi) \frac{v_m^2}{a} \frac{\sqrt{2w}}{\kappa} = M^*. \tag{32}$$

Now, the simplified governing equations are [73]

$$\frac{1}{M_\phi^{eff}}\frac{\partial \phi}{\partial t} = \kappa^2 \nabla^2 \phi - w\frac{\partial g}{\partial \phi} - \frac{1}{v_m}\frac{\partial h_\alpha}{\partial \phi}\left[f_\alpha - f_\gamma - (h_\alpha \tilde{\mu}_\alpha + h_\gamma \tilde{\mu}_\gamma)(c_\alpha - c_\gamma)\right], \tag{33}$$

$$\frac{1}{M_\phi^{eff}} = \frac{1}{M_\phi} + \left(\frac{\partial h_\alpha}{\partial \phi}\right)^2 \frac{(c_\gamma - c_\alpha)^2}{M^*}, \tag{34}$$

$$\frac{\partial c_\alpha^*}{\partial t} = \frac{M^*}{v_m} h_\gamma(\tilde{\mu}_\gamma - \tilde{\mu}_\alpha) + \frac{\partial h_\alpha}{\partial \phi}\frac{\partial \phi}{\partial t}(c_\gamma - c_\alpha), \tag{35}$$

$$\frac{\partial c_\gamma^*}{\partial t} = \frac{M^*}{v_m} h_\alpha \left(\tilde{\mu}_\alpha - \tilde{\mu}_\gamma\right) + \frac{\partial h_\alpha}{\partial \phi} \frac{\partial \phi}{\partial t}\left(c_\gamma - c_\alpha\right), \tag{36}$$

$$J = -M_c \nabla \left(h_\alpha \tilde{\mu}_\alpha + h_\gamma \tilde{\mu}_\gamma\right), \tag{37}$$

in which Eqs. (35) and (36) recover Steinbach et al.'s expressions for short-range solute redistribution [54]. In contrast, the present long-range diffusion $J$ is significantly different, which means the long range of one phase is also driven by the chemical gradient of the other. This is not strange because the separate concentration field is continuously changed by trans-sharp-interface diffusion. Moreover, we can find Eqs. (33)-(37) follow the Onsager's reciprocal relationship [29,30] as the following

$$\begin{bmatrix} \partial \phi/\partial t \\ \partial c_\alpha^*/\partial t \\ \partial c_\gamma^*/\partial t \\ J \end{bmatrix} = \begin{bmatrix} L_{\phi\phi} & L_{\phi\alpha}^* & L_{\phi\gamma}^* & 0 \\ L_{\alpha\phi}^* & L_{\alpha\alpha}^* & L_{\alpha\gamma}^* & 0 \\ L_{\gamma\phi}^* & L_{\gamma\alpha}^* & L_{\gamma\gamma}^* & 0 \\ 0 & 0 & 0 & M_c \end{bmatrix} \begin{bmatrix} \delta F/\delta \phi \\ \delta F/\delta c_\alpha \\ \delta F/\delta c_\gamma \\ v_m \nabla \left(\dfrac{\delta F}{\delta c_\alpha} + \dfrac{\delta F}{\delta c_\gamma}\right) \end{bmatrix}, \tag{38}$$

where the kinetic coefficients are

$$L_{\phi\phi} = -M_{eff}, \tag{39}$$

$$L_{\phi\alpha}^* = L_{\alpha\phi}^* = \frac{M_{eff}}{v_m} \frac{\partial h_\alpha}{\partial \phi}\left(c_\alpha - c_\gamma\right), \tag{40}$$

$$L_{\phi\gamma}^* = L_{\gamma\phi}^* = \frac{M_{eff}}{v_m} \frac{\partial h_\alpha}{\partial \phi}\left(c_\alpha - c_\gamma\right) \tag{41}$$

$$L_{\alpha\alpha}^* = \frac{M_{eff}}{v_m}\left[\frac{\partial h_\alpha}{\partial \phi}\left(c_\alpha - c_\gamma\right) - \frac{h_\gamma}{h_\alpha}\right], \tag{42}$$

$$L_{\alpha\gamma}^* = L_{\gamma\alpha}^* = \frac{M_{eff}}{v_m}\left[\frac{\partial h_\alpha}{\partial \phi}\left(c_\alpha - c_\gamma\right) + 1\right], \tag{43}$$

$$L_{\gamma\gamma}^* = \frac{M_{eff}}{v_m}\left[\frac{\partial h_\alpha}{\partial \phi}\left(c_\alpha - c_\gamma\right) - \frac{h_\alpha}{h_\gamma}\right]. \tag{44}$$

Therefore, the present model is consistent with the classical irreversible thermodynamics, i.e., MEPP

or TEP [29–33].

Now, the change rate of separated and overall concentration fields are given by

$$\frac{\partial c_\alpha}{\partial t} = \nabla v_m M_c \nabla \left( h_\alpha \tilde{\mu}_\alpha + h_\gamma \tilde{\mu}_\gamma \right) + \frac{M^*}{v_m} h_\gamma \left( \tilde{\mu}_\gamma - \tilde{\mu}_\alpha \right) + \frac{\partial h_\alpha}{\partial \phi} \frac{\partial \phi}{\partial t} \left( c_\gamma - c_\alpha \right), \tag{45}$$

$$\frac{\partial c_\gamma}{\partial t} = \nabla v_m M_c \nabla \left( h_\alpha \tilde{\mu}_\alpha + h_\gamma \tilde{\mu}_\gamma \right) + \frac{M^*}{v_m} h_\alpha \left( \tilde{\mu}_\alpha - \tilde{\mu}_\gamma \right) + \frac{\partial h_\alpha}{\partial \phi} \frac{\partial \phi}{\partial t} \left( c_\gamma - c_\alpha \right), \tag{46}$$

and

$$\frac{\partial c}{\partial t} = \nabla v_m M_c \nabla \left( h_\alpha \tilde{\mu}_\alpha + h_\gamma \tilde{\mu}_\gamma \right). \tag{47}$$

It is evident that only the long-range term remain in the Eq. (47), meaning that the overall $c$ is still conserved as Eq. (5).

Before proceeding, some numerical features of this model should be emphasized. Due to the obviation of equal diffusion potential, there is no extra work to solve $c_\alpha$ and $c_\gamma$ with an additional couped set of non-linear equations, which is computationally efficient as the similar models [54,72]. Moreover, the current long-range diffusion [Eq. (47)] is determined by the interpolated diffusion potential. At the bulk region of $\alpha$, the influence of $\gamma$ is naturally eliminated by $h_\gamma = 0$ (vice versa). Therefore, as the Kim-Kim-Suzuki (KKS) model [53], the overall kinetics is not affected by the concentration of one phase at the other bulk. In contrast, the long-range diffusion of Steinbach et al.'s previous model [54] is governed by the diffusion potential of individual phases. On the one hand, it will result in ill-defined regime for $\phi = 0, 1$, which need an artificially "cut off" of interface and bulks during the simulations [54], or setting a threshold value of $\phi$, e.g., $\phi_{max} = 0.9999$ and $\phi_{min} = 0.0001$ [74]. On the other hand, if employing the "cut-off" method, the responses of interface kinetics will be non-unique, depending on the boundary condition of one phase's concentration at the other.

## III. Thermodynamics of Interface

### A. Equilibrium properties

Before analyzing the non-equilibrium thermodynamics of the diffuse interface, we first briefly introduce the equilibrium state of this model, which is the same as the Kim-Kim-Suzuki (KKS) model [53], i.e., the contributions of interface and bulks are decoupled. The equilibrium phase-field profile is given by

$$\phi_{eq}(x) = \frac{1}{2}\left[1 - \tanh\left(\frac{\sqrt{w}}{\sqrt{2}\kappa} x\right)\right], \tag{48}$$

whose gradient has been used in deriving the $M^*$ for trans-sharp-interface diffusion, i.e., Eq. (32). Then, the interface energy $\sigma_{\alpha\gamma}$ and interface thickness $2\delta$ are [53]

$$\sigma_{\alpha\gamma} = \kappa^2 \int_{-\infty}^{+\infty} \left(\frac{d\phi_{eq}}{dx}\right)^2 dx = \frac{\kappa\sqrt{w}}{3\sqrt{2}}, \tag{49}$$

$$2\delta = \chi\sqrt{2}\frac{\kappa}{\sqrt{w}}, \tag{50}$$

in which $\chi = 2.2$ for defining $2\delta$ from $\phi = 0.1$ to $\phi = 0.9$, and $\chi = 2.94$ for $\phi = 0.05$ to $\phi = 0.95$. Note that submitting Eq. (50) into the Eq. (32), $M^*$ will be scaled with $1/(2a\delta)$ as Steinbach et al.'s work [54]. It indicates that the present derivation based on the comparison with classical sharp interface models [24–27] is consistent with previously microscopic consideration of atomic jumps.

### B. Non-equilibrium sharp interface in RVEs

Due to consistency between Eqs. (17), (35), (36) and Eqs. (A4), (A7), (A8), there should be a similar non-equilibrium thermodynamics of RVEs as the classical sharp interface models [19–28]. However, no unified cognition about the free energy dissipation by different processes has been reached in the classical sharp interface model, e.g., the discrepancy between Hillert et al.'s one-step [19–21,75] and Wang et al.'s two-step concepts [24,26,27]. In this section, we will inherit some viewpoints of their works and generalize them into a unified "two-part" mechanism.

Although existing disagreement in understanding the energy dissipation of different processes, there is the same expression for the total driving force of sharp interface, i.e., Eq. (A12). As for this model, we can reorganize Eq. (17) as the following

$$\frac{\partial c_\alpha^*}{\partial t} = \frac{1}{h_\alpha} \frac{\partial h_\alpha}{\partial \phi} \frac{\partial \phi}{\partial t} \left( c_{trans}^{SI} - c_\alpha \right), \tag{51}$$

and

$$\frac{\partial c_\gamma^*}{\partial t} = \frac{1}{h_\gamma} \frac{\partial h_\gamma}{\partial \phi} \frac{\partial \phi}{\partial t} \left( c_{trans}^{SI} - c_\gamma \right), \tag{52}$$

where

$$c_{trans}^{SI} = c_\alpha + h_\alpha \frac{\partial c_\alpha^*}{\partial t} \frac{\partial t}{\partial h_\alpha} = c_\gamma + h_\gamma \frac{\partial c_\gamma^*}{\partial t} \frac{\partial t}{\partial h_\gamma}, \tag{53}$$

which are similar to Eq. (A11) of the classical sharp interface models. $c_{trans}^{SI}$ is the composition of materials transferred across the interface, which is defined in a frame of reference fixed to the migrating sharp interface [21].

According to the change rate of total free energy [Eq. (8)], the change rate of sharp interfaces in RVEs is given by

$$\dot{F}_{RVE}^{SI} = \int_\Omega \left\{ \frac{\partial \phi}{\partial t} \frac{\delta F}{\delta \phi} + \frac{\partial c_\alpha^*}{\partial t} \frac{\delta F}{\delta c_\alpha} + \frac{\partial c_\gamma^*}{\partial t} \frac{\delta F}{\delta c_\gamma} \right\} d\Omega. \tag{54}$$

Submitting Eqs. (51)-(53) in Eq. (54) and dividing the change rate of phase-field, we then obtain the total driving force for sharp interface in RVEs,

$$\Delta f_{tot}^{SI} = f_\alpha - f_\gamma + \tilde{\mu}_\alpha \left( c_{trans}^{SI} - c_\alpha \right) + \tilde{\mu}_\gamma \left( c_\gamma - c_{trans}^{SI} \right), \tag{55}$$

which is the same as the classical sharp interface models, i.e., Eq. (A12). Only in the case of equal diffusion potential, $\Delta f_{tot}^{SI}$ recovers the grand potential difference between $\alpha$ and $\gamma$ [54]

$$\Delta f_{tot}^{SI}\left(\tilde{\mu}_\alpha = \tilde{\mu}_\gamma\right) = f_\alpha - c_\alpha \tilde{\mu}_\alpha - \left(f_\gamma - c_\gamma \tilde{\mu}_\gamma\right). \tag{56}$$

Eq. (55) indicates that the total driving force for sharp phase transformation is consumed to drive a diffusional transformation from $f_\gamma(c_\gamma)$ to $f_\alpha(c_\alpha)$ and two trans-sharp-interface diffusion fluxes, i.e., Eqs. (51) and (52), for which " $f_\alpha - f_\gamma$ ", " $\tilde{\mu}_\gamma$ ", and " $\tilde{\mu}_\alpha$ " are their conjugated driving forces given by Onsager's reciprocal relationship in Eq. (38). This is Wang and Kuang et al.'s two-step concept [24,26,27], whose graphical construction of free energy dissipation is given in Figure. 2(a). However, to distinctly understand different dissipative processes and their thermodynamics, it is necessary to find independent fluxes and forces.

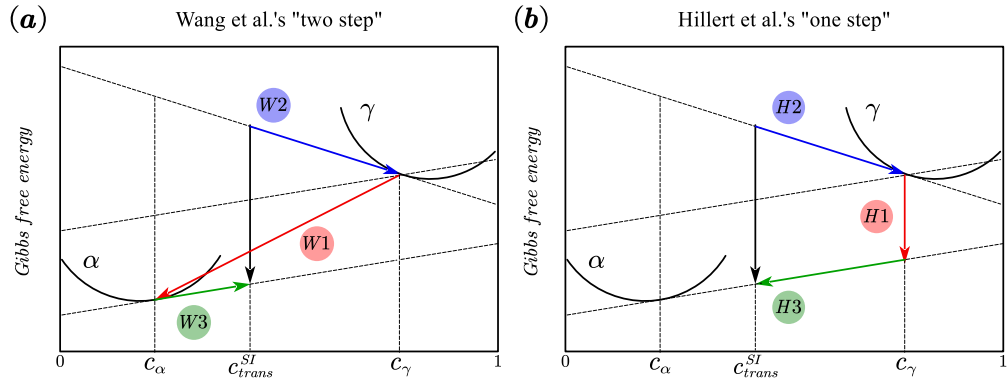

FIG. 2. Graphical constructions of free energy dissipation for interface migration and trans-sharp-interface diffusions in (a) Wang et al.'s two-step concept [24,26,27] and (b) Hillert et al.'s one-step concept [19–21,75]. In Wang et al.'s understanding (a): the interface migration "$W1$" and trans-interface-diffusion "$W2$" and "$W3$" are dependent processes occurring simultaneously, in which $W1 = f_\alpha - f_\gamma$, $W2 = \tilde{\mu}_\gamma\left(c_\gamma - c_{trans}^{SI}\right)$, and $W3 = \tilde{\mu}_\alpha\left(c_{trans}^{SI} - c_\alpha\right)$. Because the diffusion fluxes of $W2$ and $W3$ are different, they called this picture as the trans-interface diffusion in two-step. In Hillert et al.'s model (b): the interface migration "$H1$" and trans-interface-diffusion "$H2$" and "$H3$" are independent, where $H1 = f_\alpha - f_\gamma - \tilde{\mu}_\alpha\left(c_\alpha - c_\gamma\right)$, $H2 = \tilde{\mu}_\gamma\left(c_\gamma - c_{trans}^{SI}\right)$, and $H3 = -\tilde{\mu}_\alpha\left(c_\gamma - c_{trans}^{SI}\right)$. Since the equal diffusion fluxes of $H2$ and $H3$, they can be combined into one-step trans-interface diffusion, i.e., $H2 + H3 = \left(\tilde{\mu}_\gamma - \tilde{\mu}_\alpha\right)\left(c_\gamma - c_{trans}^{SI}\right)$.

Using Eqs. (33), (34), submitting Eqs. (51), (52) into Eqs. (35), (36), there are three independent fluxes with three independent driving forces, yielding as

$$\frac{1}{M_\phi^{eff}}\frac{\partial \phi}{\partial t} = \kappa^2 \nabla^2 \phi - w\frac{\partial g}{\partial \phi} - \frac{1}{v_m}\frac{\partial h_\alpha}{\partial \phi}\left[f_\alpha - f_\gamma - \left(h_\alpha \tilde{\mu}_\alpha + h_\gamma \tilde{\mu}_\gamma\right)\left(c_\alpha - c_\gamma\right)\right]; \tag{57}$$

$$\frac{1}{M_\phi^{eff}} = \frac{1}{M_\phi} + \left(\frac{\partial h_\alpha}{\partial \phi}\right)^2 \frac{\left(c_\gamma - c_\alpha\right)^2}{M^*}, \tag{58}$$

and

$$\frac{1}{M_{\alpha,eff}^*}\frac{\partial c_\alpha^*}{\partial t} = -\frac{h_\alpha h_\gamma}{v_m}\left(\tilde{\mu}_\alpha - \tilde{\mu}_\gamma\right); \tag{59}$$

$$M_{\alpha,eff}^* = \frac{M^*}{h_\alpha}\frac{c_{trans}^{SI} - c_\alpha}{c_{trans}^{SI} - \left(h_\alpha c_\gamma + h_\gamma c_\alpha\right)}, \tag{60}$$

and

$$\frac{1}{M_{\gamma,eff}^*}\frac{\partial c_\gamma^*}{\partial t} = -\frac{h_\alpha h_\gamma}{v_m}\left(\tilde{\mu}_\alpha - \tilde{\mu}_\gamma\right); \tag{61}$$

$$M_{\gamma,eff}^* = \frac{M^*}{h_\gamma}\frac{c_{trans}^{SI} - c_\gamma}{c_{trans}^{SI} - \left(h_\alpha c_\gamma + h_\gamma c_\alpha\right)}. \tag{62}$$

Note that $h_\alpha c_\gamma + h_\gamma c_\alpha$ is not the overall concentration. The change rate of total free energy with RVEs can then be reproduced by

$$\begin{aligned}-\frac{1}{v_m}\frac{\partial h_\alpha}{\partial \phi}\frac{\partial \phi}{\partial t}\left[\Delta f_{tot}^{SI}\right] &= \frac{1}{M_\phi^{eff}}\left(\frac{\partial \phi}{\partial t}\right)^2 + \frac{1}{M_{\alpha,eff}^*}\left(\frac{\partial c_\alpha^*}{\partial t}\right)^2 + \frac{1}{M_{\gamma,eff}^*}\left(\frac{\partial c_\gamma^*}{\partial t}\right)^2 \\ &= \frac{1}{M_\phi}\left(\frac{\partial \phi}{\partial t}\right)^2 + \frac{1}{M_\alpha^*(\phi)}\left(\frac{\partial c_\alpha^*}{\partial t}\right)^2 + \frac{1}{M_\gamma^*(\phi)}\left(\frac{\partial c_\gamma^*}{\partial t}\right)^2,\end{aligned} \tag{63}$$

where the right hand of Eq. (63) is the free energy dissipation of RVEs. It is shown the total dissipation by independent fluxes and driving forces are the same as that of dependent fluxes and driving forces.

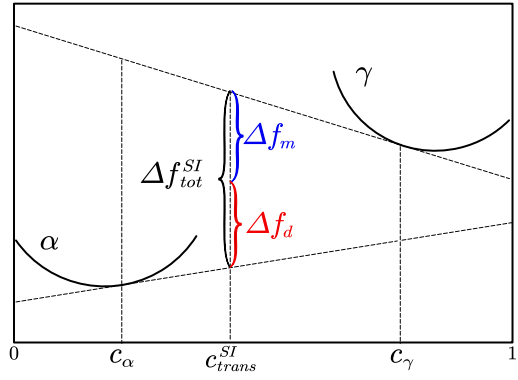

FIG. 3. Partition of total driving force $\Delta f_{tot}^{SI}$ by interface migration $\Delta f_m$ and trans-sharp-interface diffusion $\Delta f_d$ for the present sharp interface in RVEs.

Now, we have the balance of driving forces and energy dissipation for sharp interface in RVEs, i.e., $\Delta f_m$ for sharp interface migration and $\Delta f_d$ for trans-sharp-interface diffusion

$$\Delta f_m = f_\alpha - f_\gamma - \left(h_\alpha \tilde{\mu}_\alpha + h_\gamma \tilde{\mu}_\gamma\right)\left(c_\alpha - c_\gamma\right), \tag{64}$$

$$\Delta f_d = \left(\tilde{\mu}_\alpha - \tilde{\mu}_\gamma\right)\left[c_{trans}^{SI} - \left(h_\alpha c_\gamma + h_\gamma c_\alpha\right)\right], \tag{65}$$

whose graphical construction of free energy changes is exhibited in Figure. 3. Comparing Figures. 2 and 3 indicates that the present understanding differs from all existing theories.

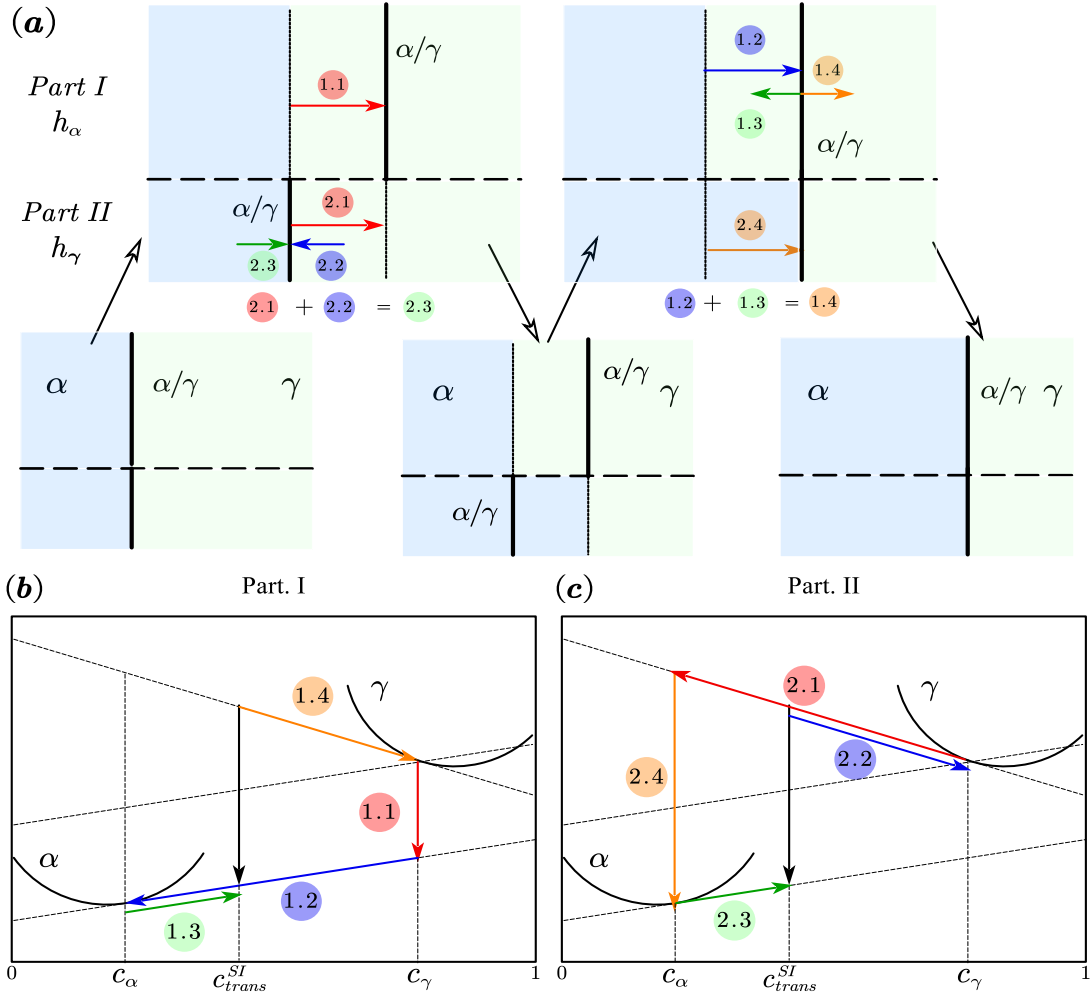

FIG. 4. (a): The "two-part" schematics for "sharp-interface" type phase transformation in RVEs; (b) and (c): Graphical constructions of molar free energy changes for Parts I and II, respectively. Parts I and II are divided by the dashed lines in (a). In each part, the phase boundary is represented by the black solid line, while the background colors represent the lattice concentrations of $\alpha$ and $\gamma$. The (1.1-1.4) and (2.1-2.4) denote the kinetic processes of Part I and II, respectively.

Both Eq. (64) and Eq. (65) are averaged by phase fractions $h_\alpha$ and $h_\gamma$, thus, we may further rewritten them as

$$\Delta f_m^I = h_\alpha \left[ f_\alpha - f_\gamma - \tilde{\mu}_\alpha \left( c_\alpha - c_\gamma \right) \right], \tag{66}$$

$$\Delta f_d^I = h_\alpha \left( \tilde{\mu}_\alpha - \tilde{\mu}_\gamma \right) \left( c_{trans}^{SI} - c_\gamma \right), \tag{67}$$

and

$$\Delta f_m^{II} = h_\gamma \left[ f_\alpha - f_\gamma - \tilde{\mu}_\gamma \left( c_\alpha - c_\gamma \right) \right], \tag{68}$$

$$\Delta f_d^{II} = h_\gamma \left( \tilde{\mu}_\alpha - \tilde{\mu}_\gamma \right)\left( c_{trans}^{SI} - c_\alpha \right). \tag{69}$$

It is evident to see the present Eqs. (66) and (67) are the same as Hillert et al.'s one-step viewpoint of energy dissipation [19–21,75], as given by $H1$ and $H2+H3$ in Figure. 2(b). This means that this sharp interface transformation follows a "two-part" mechanism, i.e., Part I ($h_\alpha$) and Part II ($h_\gamma$).

We first discuss the phase transformation mechanisms of Part I with the same scenario as Hillert et al.'s works [19–21,75]. As shown by (1.1) in Figure. 4(a), the phase boundary first transforms from $\gamma(c_\gamma)$ to $\alpha(c_\gamma)$ in a diffusionless way, whose corresponding change in free energy is shown in Figure. 4(b), i.e., Eq. (66). Since concentration remains the same in (1.1), two trans-sharp-interface diffusions in Eqs. (35) and (36) are offset. The net dissipation of these two fluxes also equals 0 according to Eqs. (51) and (52). After that, the metastable $\alpha(c_\gamma)$ transforms into $\alpha(c_\alpha)$ by desorbing atoms as (1.2) in Figure. 4(a), i.e., $c_\alpha - c_\gamma$. The dissipation of $c_\alpha - c_\gamma$ in $\alpha$ (driven by $\tilde{\mu}_\alpha$) is then denoted by

$$\Delta f_d^I \Big|_{1.2} = h_\alpha \tilde{\mu}_\alpha \left( c_\alpha - c_\gamma \right). \tag{70}$$

Simultaneously, the composition changes by (1.2) will induce a moving tendency of the phase boundary. The concentration and phase boundary fluctuations will cause two non-compensated trans-sharp-interface fluxes, i.e., (1.3) and (1.4) in Figure. 4(a). They are also consistent with Eqs. (51) and (52), respectively. The corresponding dissipation of $c_{trans}^{SI} - c_\alpha$ in $\alpha$ and $c_\gamma - c_{trans}^{SI}$ in $\gamma$ are then represented by

$$\Delta f_d^I \Big|_{1.3} = h_\alpha \tilde{\mu}_\alpha \left( c_{trans}^{SI} - c_\alpha \right), \tag{71}$$

$$\Delta f_d^I \Big|_{1.4} = h_\alpha \tilde{\mu}_\gamma \left( c_\gamma - c_{trans}^{SI} \right), \tag{72}$$

which are the same as Wang et al.'s $W2$ and $W3$ in Figure. 2(a). Summarizing Eqs. (70)-(72) will recover the whole dissipation of trans-sharp-interface diffusion in Part. I, Eq. (67). Note that (1.3) and (1.4) will induce an opposite moving tendency of the phase boundary, and the sharp interface is finally not moved according to Eq. (17). In other words, the net material transferred from $\alpha$ to interface equals that transferred from the interface to $\gamma$, i.e., $(1.2)+(1.3)=(1.4)=c_\gamma - c_{trans}^{SI}$. Hence, there are no retained atoms moving the phase boundary.

It should be mentioned that all former analyses are based on a number fixed frame of reference. The compositions of growing $\alpha$ and parent $\gamma$ are still $c_\alpha$ and $c_\gamma$, respectively, for which the tangent lines of free energy in Figure 4(b) and 4(c) are drawn from $c_\alpha$ and $c_\gamma$. In contrast, defined by the net flux across the interface, $c_{trans}^{SI}$ is the material composition in a frame of reference fixed to the migrating interface [21]. It is applied to represent two diffusion fluxes, Eqs. (51) and (52) always exist unless the phase transformation stops. Alternatively, if one continuously views Part I of Figure. 4(a) at different times, there always be phase transformation from $\gamma(c_\gamma)$ to $\alpha(c_\alpha)$ and two fluxes being transferred. This is Wang et al.' understanding inspired from the total driving force Eq. (55).

The processes in Part II are similar to Part I. The three trans-sharp-interface diffusions, (1.1)-(1.3) first transform $\gamma(c_\gamma)$ to metastable $\gamma(c_\alpha)$, whose corresponding changes in free energy are

$$\Delta f_d^{II}\big|_{2.1} = h_\gamma \tilde{\mu}_\gamma (c_\alpha - c_\gamma), \tag{73}$$

$$\Delta f_d^{II}\big|_{2.2} = h_\gamma \tilde{\mu}_\gamma (c_\gamma - c_{trans}^{SI}), \tag{74}$$

$$\Delta f_d^{II}\big|_{2.3} = h_\gamma \tilde{\mu}_\alpha (c_{trans}^{SI} - c_\alpha). \tag{75}$$

Then, the phase boundary migration (2.4) makes the metastable $\gamma(c_\alpha)$ diffusionless transform into $\alpha(c_\alpha)$, whose free energy dissipation has been given in Eq. (68). The present diffusion-assisted displacive mechanism is similar to a recent experimental finding of Mg-Al alloy [61] by neutron total scattering. It was found that the short-range atomic redistributions first change the compositions of short-range ordered cells (or groups), and then phase transformations proceed in a diffusionless manner.

Additionally, the classical sharp interface models can also be reinterpreted by the present "two-part" understanding, as illustrated in Appendix.

**C. Non-equilibrium integrated diffuse interface**

With a distinct understanding of non-equilibrium thermodynamics in each RVE of the diffuse interface, it is now necessary to explore the relationship between the present PF model with the classical diffuse interface theories by Cahn [56] and Hillert and Sundman [57]. The present model is analytically tractable by employing the standard coordinate transformation $\partial/\partial t = -V_n \partial/\partial x$ for 1-D steady state, i.e.,

$$\int_{-\delta}^{+\delta}\left\{\frac{V_n}{M_\phi}\left(\frac{\partial \phi}{\partial x}\right)^2\right\}dx = \int_{-\delta}^{+\delta}\left\{\frac{\delta F}{\delta \phi}\frac{\partial \phi}{\partial x} + \frac{\delta F_{ad}}{\delta \phi}\frac{\partial \phi}{\partial x}\right\}dx, \tag{76}$$

$$\int_{-\delta}^{+\delta}\left\{\frac{V_n}{M_\alpha(\phi)}\left(\frac{\partial c_\alpha^*}{\partial x}\right)^2\right\}dx = \int_{-\delta}^{+\delta}\left\{\frac{\delta F}{\delta c_\alpha}\frac{\partial c_\alpha^*}{\partial x} + \frac{\delta F_{ad}}{\delta c_\alpha}\frac{\partial c_\alpha^*}{\partial x}\right\}dx, \tag{77}$$

$$\int_{-\delta}^{+\delta}\left\{\frac{V_n}{M_\gamma(\phi)}\left(\frac{\partial c_\gamma^*}{\partial x}\right)^2\right\}dx = \int_{-\delta}^{+\delta}\left\{\frac{\delta F}{\delta c_\gamma}\frac{\partial c_\gamma^*}{\partial x} + \frac{\delta F_{ad}}{\delta c_\gamma}\frac{\partial c_\gamma^*}{\partial x}\right\}dx. \tag{78}$$

In addition, the steady trans-diffuse-interface diffusion flux according to Eq. (4) is given by

$$J = \frac{V_n}{v_m}\left(c - c_{trans}^{DI}\right) = \frac{V_n}{v_m}\left(c_{i=\alpha,\gamma} - c_{trans}^{DI}\right), \tag{79}$$

where $c_{trans}^{DI}$ is the concentration at a reference point where the long-range diffusion $J\left(c_{trans}^{DI}\right) = 0$. Combining Eqs. (79), (37) and integrating in part, we have

$$\int_{-\delta}^{+\delta}\left\{\frac{1}{V_n}\frac{J^2}{M_c(\phi)}\right\}dx = \int_{-\delta}^{+\delta}\left\{\frac{\delta F}{\delta c_\alpha}\frac{\partial \hat{c}_\alpha}{\partial x} + \frac{\delta F}{\delta c_\gamma}\frac{\partial \hat{c}_\gamma}{\partial x}\right\}dx \\ + \tilde{\mu}_\gamma^\delta\left(c_{trans}^{DI} - c_\gamma^\delta\right) - \tilde{\mu}_\alpha^\delta\left(c_{trans}^{DI} - c_\alpha^{-\delta}\right). \tag{80}$$

Summarizing Eqs. (76)-(78) and (80) provides the balance of total driving force $\Delta f_{tot}^{DI}$ and free energy dissipation $Q_{tot}$ for the diffuse interface, i.e.,

23 / 52

$$\Delta f_{tot}^{DI} = Q_{tot}, \tag{81}$$

$$Q_{tot} = \int_{-\delta}^{+\delta} \left\{ \frac{V}{M_\phi} \left( \frac{\partial \phi}{\partial x} \right)^2 + \sum_{i=\alpha,\gamma} \frac{V_n}{M_i^*} \left( \frac{\partial c_i^*}{\partial x} \right)^2 + \frac{1}{V_n} \frac{J^2}{M_c} \right\} dx, \tag{82}$$

$$\Delta f_{tot}^{DI} = f_\gamma^\delta - f_\alpha^{-\delta} + \tilde{\mu}_\gamma^\delta \left( c_{trans}^{DI} - c_\gamma^\delta \right) - \tilde{\mu}_\alpha^\delta \left( c_{trans}^{DI} - c_\alpha^{-\delta} \right). \tag{83}$$

The quadratic form of Eq. (82) is consistent with the initial assumption of free energy dissipation in the linear thermodynamics, Eq. (10). Also, it proves that the whole diffuse interface is dissipative whose entropy is increasing. As for the driving force $\Delta f_{tot}^{DI}$, Eq. (83) is found to have the similar expression as the sharp interface, i.e., $\Delta f_{tot}^{SI}$ of Eq. (55). Hence, the reference concentration $c_{trans}^{DI}$ may be defined as the composition of materials transferred across the diffuse interface, and $c_{trans}^{DI} = c_\alpha^{-\delta}$ if the diffusion in growing phase is negligible, e.g., solidification.

Using Eqs. (37) and (79), the last term in Eq. (83) can be transformed into its equivalent forms,

$$\begin{aligned} Q_d &= \int_{-\delta}^{+\delta} \left\{ \frac{1}{V_n} \frac{J^2}{M_c} \right\} dx \\ &= \int_{-\delta}^{+\delta} \left\{ \frac{V_n}{v_m^2} \frac{\left( c - c_{trans}^{DI} \right)^2}{M_c} \right\} dx \\ &= -\int_{-\delta}^{+\delta} \left\{ \frac{1}{v_m} \left( c - c_{trans}^{DI} \right) \nabla \left( h_\alpha \tilde{\mu}_\alpha + h_\gamma \tilde{\mu}_\gamma \right) \right\} dx. \end{aligned} \tag{84}$$

If neglecting the diffusion in growing phase, it is evident that the third row recovers the solute drag force of the classical diffuse interface theories by Cahn [56] and Hillert and Sundman [57].

However, the relationship between the former three terms and the classical interface friction is unclear. If assuming the effective phase-field mobility $M_{eff}$ of Eq. (34) as a constant, Eq. (82) can then be reorganized by Eqs. (17), (35) and (36) as the following

$$Q_{tot} = Q_m + Q_d + Q_e,$$
$$Q_m = \frac{V_n}{\left(\kappa^2/\sigma_{\alpha\gamma}\right)M_\phi^{eff}},$$
$$Q_e = \int_{-\delta}^{+\delta}\left\{\frac{V_n}{M^* v_m} h_\alpha h_\gamma \left(\tilde{\mu}_\alpha - \tilde{\mu}_\gamma\right)^2\right\}dx. \tag{85}$$

Now, we can see the first term recovers the classical interface friction, which is usually proportionate to the interface velocity $V_n$. Note that part of dissipation by $\partial c_i/\partial t$ has been incorporated into $Q_m$ by $M_\phi^{eff}$. Then, determined by diffusion potential difference, the remained $Q_e$ still somewhat reflects the influence of non-equilibrium state in the diffuse interface. In the case of near-equilibrium state, Eq. (85) naturally reproduces the classical diffuse interface theories,

$$Q_{tot} \approx \frac{V_n}{\left(\kappa^2/\sigma_{\alpha\gamma}\right)M_\phi^{eff}} + \int_{-\delta}^{+\delta}\left\{\frac{1}{V_n}\frac{J^2}{M_c}\right\}dx. \tag{86}$$

Therefore, the $M_{eff}$ is expressed as

$$M_\phi^{eff} = m\frac{\sigma_{\alpha\gamma}}{\kappa^2}. \tag{87}$$

where the intrinsic interface mobility $m$ is determined experimentally.

## IV.  Non-equilibrium Interface Kinetics

To validate the present phase-field model in representing non-equilibrium interface kinetics, we apply it to the 1-D rapidly solidified Al-Cu alloys compared with Haapalehto et al.'s molecular-dynamic (MD) simulations [16]. Having the same thickness as the MD simulations, the current diffuse interface is expected to represent the main characteristics of the actual solid-liquid diffuse interface. The comparison with MD will mainly focus on the solute trapping and solute drag, especially the partial drag phenomena at the low-velocity regime and complete trapping at the high-velocity regime.

## A. Thermophysical parameters

To precisely compare with Haapalehto et al.'s molecular-dynamic (MD) simulations [16], we directly adopt the MD simulated phase diagram, although it overpredicts the actual melting point and liquidus slope [16]. The dilute approximations of thermodynamics based on a linear phase diagram are as follows:

$$f_l = RT\{c_l \ln(c_l) + (1-c_l)\ln(1-c_l)\}, \tag{88}$$

$$f_l = RT\left\{c_s \ln(c_s) + (1-c_s)\ln(1-c_s) - c_s \ln(k_e) + (1-c_s)\ln\left[\frac{1-c_l^{eq}}{1-c_s^{eq}}\right]\right\}, \tag{89}$$

$$c_s^{eq} = k_e \frac{T-T_m}{m_e}; \quad c_l^{eq} = \frac{T-T_m}{m_e} \tag{90}$$

where $R$ is the ideal gas constant and $T$ is the solidification temperature. Other thermodynamic parameters in Eqs. (88)-(90) are given in Table. 1. Figure. 5 displays the comparison of equilibrium concentrations predicted by Eq. (90) and the MD simulations, in which there is a good agreement.

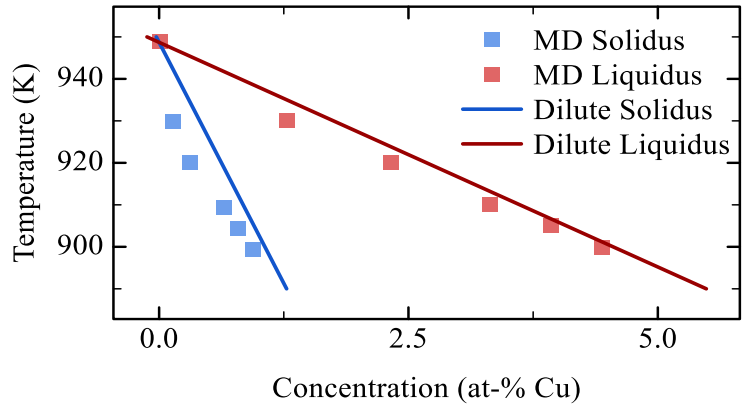

FIG. 5. Al-Cu linear phase diagram predicted by Eqs. (88) and (89), compared with the MD simulations [16].

Additionally, the diffusion coefficient $D_l$ in liquid is the same as the MD, while the solid diffusion coefficient is set to be significantly smaller, i.e., $D_l/1000$. The intrinsic mobility $m$ for solid-liquid interface is generally expressed as

$$m = \frac{v_m V_0}{RT}, \tag{91}$$

where $V_0$ is the maximal possible transformation velocity. Its value is usually on the order (but less than) the sound velocity in liquid, $\sim 10^3$ m s$^{-1}$. In this work, $V_0$ is treated as a fitting parameters since two nominal compositions (1 at.% and 2 at.%) are simulated, and $V_0$ may have a strong composition dependence [76].

Tab. 1. Thermophysical parameters of Al-Cu alloy computed and fitted from MD simulations [16].

| Parameters | | Unit | Value |
|---|---|---|---|
| Interface thickness | $2\delta$ | m | $2\times 10^{-9}$: $\phi$=0.05:0.95 |
| Interface energy | $\sigma_{\alpha\gamma}$ | J m$^{-2}$ | 0.127 |
| Diffusivity in liquid | $D_l$ | m$^2$ s$^{-1}$ | $2.77\times 10^{-7}\exp(-9.64\times 10^4/RT)$ |
| Diffusivity in solid | $D_s$ | m$^2$ s$^{-1}$ | $D_l/1000$ |
| Melting point of Cu | $T_m$ | K | 948.7 |
| Liquidus slope | $m_e$ | K at.$^{-1}$ | $-1070$ |
| Equilibrium partition coefficient | $k_e$ | – | 0.233 |
| Interatomic spacing | $a$ | m | $2\times 10^{-10}$ |
| Maximal possible velocity 1at.% | $V_0$ | m s$^{-1}$ | 1200 |
| Maximal possible velocity 2at.% | $V_0$ | m s$^{-1}$ | 900 |
| Critical velocity for $k(V)=1$ | $V_c$ | m s$^{-1}$ | 12.5 |
| Atomic volume | $v_m$ | m$^3$ mol$^{-1}$ | $1\times 10^{-5}$ |

## B. Long-diffusion with relaxation effect

In rapid solidification, the velocity of the solid-liquid interface can approach or exceed the characteristic diffusion velocity [$V_c$] in the liquid [24], i.e., the time for crystallization of liquid is comparable to the relaxation time of long-range diffusion $\tau_D = D_l/V_c^2$. Then, different from the classical Fick's law, the long-range flux $J$ is co-governed by the instantaneous chemical gradient and the local history of the diffusion process, which can be described by the Maxwell-Cattaneo equation

$$J + \tau_D \frac{\partial J}{\partial t} = -\frac{D_l}{v_m} \nabla c. \tag{92}$$

Application of Eq. (92) in classical sharp interface models have successfully reproduce the sharp transition from diffusional solidification to diffusionless solidification at finite velocity $V_c$, see Sobolev [77–81], Galenko [82–85], and Wang et al. [24,25].

As for the phase-field modeling, the relaxation effect was firstly introduced by Galenko and Jou [86] through a kinetic energy of diffusion flux, in which a second order time derivative $\tau_D \partial^2 c / \partial t^2$ was added in the Cahn-Hilliard equation [68]. However, when interface velocity $V_n \geq V_c$, the interfacial concentration profile is still inhomogeneous, which results in a weak (but not disappearing) solute drag effect. This is inconsistent with the atomic simulation results. To address this problem, Wang and Galenko et al. [87] then introduced the diffusion relaxation through an effective mobility,

$$J = -\left[1 + \frac{v_m}{V_c^2} \frac{1}{\nabla c} \frac{\partial J}{\partial t}\right] \frac{D_l}{v_m} \left(\frac{\partial \tilde{\mu}}{\partial c}\right)^{-1} \nabla \tilde{\mu}, \tag{93}$$

in which the [ ] is the effective coefficient. In this work, we add the same modification on $M_c$ in Eq. (37) [or (47)], i.e.,

$$M_{c,eff} = \left[1 + \frac{v_m}{V_c^2} \frac{1}{\nabla c} \frac{\partial J}{\partial t}\right] M_c(\phi). \tag{94}$$

Note that the overall concentration $c$ equals separately conserved fields $\hat{c}_\alpha$ and $\hat{c}_\gamma$. At the 1-D steady state, the concentration gradient is given by

$$\frac{\partial c}{\partial x} = \begin{cases} -\frac{1}{V_n} \frac{\partial}{\partial x}\left[v_m M_c(\phi)\left(1 - \frac{V_n^2}{V_c^2}\right) \frac{\partial}{\partial x}\left(h_\alpha \tilde{\mu}_\alpha + h_\gamma \tilde{\mu}_\gamma\right)\right], & V < V_c \\ 0, & V \geq V_c \end{cases} \tag{95}$$

which can also reproduce the homogeneous concentration profile after reaching $V_c$.

However, different from Galenko and Jou's hyperbolic equation [86] with a simple addition of $\tau_D \partial^2 c/\partial t^2$, Eqs. (47) and (94) are numerically tricky to be solved when simulating the transient solidification. In the low-velocity regime, it has been found that the interface temperature predicted by transient solidification is closer to reality than the steady assumption. Thus, with the aim of investigating transient interface kinetics, we change the effective coefficient as the following

$$M_{c,eff} = \left[1 + \frac{v_m}{V_c^2} \frac{1}{\nabla c} \frac{\partial J}{\partial t}\right] M_l, \ \phi = 0, \ \text{liquid} \tag{96}$$

$$M_{c,eff} = \left[1 + \frac{1}{V_c^2} \left(\frac{\dot{\phi}}{\nabla \phi}\right)^2\right] M_c(\phi), \ 0 < \phi < 1, \ \text{interface} \tag{97}$$

$$M_{c,eff} = M_s, \ \phi = 1, \ \text{solid} \tag{98}$$

It means the bulk regions ($\phi = 1$ or $0$) and interfaces ($0 < \phi < 1$) are separated as a compromise of numerical stability. Using the simulated $\partial \phi/\partial t$ and $\nabla \phi$, Eqs. (47) for $0 < \phi < 1$ is still a parabolic equation, and liquid bulk region is then solved by a standard hyperbolic equation with $\tau_D \partial^2 c/\partial t^2$. Note that the solid relaxation time $\tau_D = D_s/V_c^2$ is negligible compared with the crystallization time.

## C. Solute trapping and solute drag

Based on the former thermophysical parameters and effective mobilities for long-range diffusion, this section performs the 1-D simulations of transient solidification for Al-1 at. % Cu and Al-2 at.% Cu systems. Before proceeding, it is worthwhile to emphasize the definition of non-equilibrium partition coefficient $k(V)$.

Historically, there are various definitions of $k(V)$ for diffuse interface. For instance, Wheeler et al. [88], Ahmad et al. [89] and Kavousi et al. [90] assumed the far-field concentration as the solid concentration and the maximum concentration as the liquid phase, i.e., $k(V) = c_{\phi=0.001}/\max(c)$ [$\phi = 0$ for solid]. Based on the Danilov and Nestler's work [91], Galenko et al. [86] decomposed overall concentration $c$ into $\langle c_s \rangle + \langle c_l \rangle$ with a special function and then $k(V) = \max\langle c_s \rangle/\max\langle c_l \rangle$.

Pinomaa and Provatas [92] projected the concentration profile into a middle effective sharp interface then obtained $k(V)$. Similar to the classical diffuse interface theory [93], Lebedev et al. [94] and Wang et al. [58] defined $k(V)$ by the boundary concentrations of the diffuse interface, i.e., $k(V) = c_{\phi\to 0}/c_{\phi\to 1}$ [$\phi = 0$ for solid]. All these definitions are based on the overall concentration $c$. In contrast, Steinbach and Zhang et al. [54,95] directly define $k(V)$ by the separated $c_s$ and $c_l$, i.e., $k(V) = c_s^{\phi=0.9999}/c_l^{\phi=0.9999}$ [$\phi = 1$ for solid].

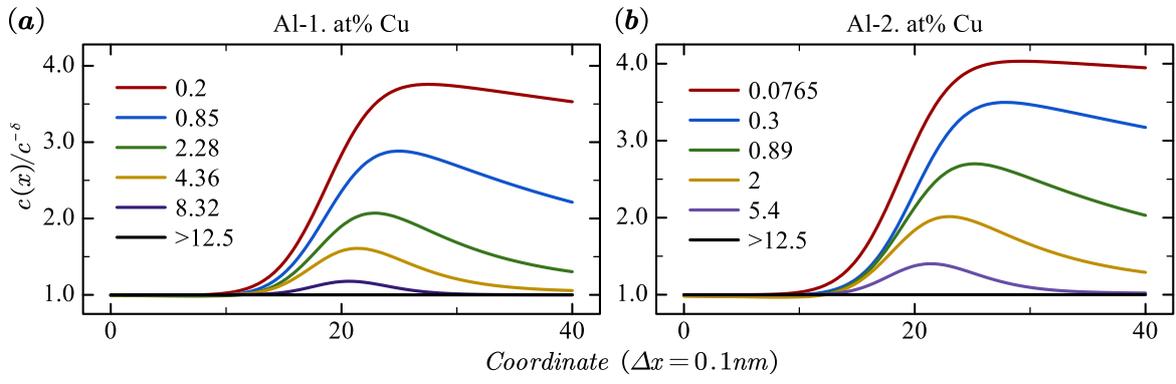

FIG. 6. Interfacial solute profiles of (a) Al-1at.% Cu and (b) Al-2at.% Cu at different velocities predicted by Eqs. (33)-(37) and (96)-(98).

Although related to the deviation from the equilibrium thermodynamics, we should note that the term "solute trapping" directly indicates the composition inhomogeneity at the interface. Thus, the definition of $k(V)$ should be consistent with experiments or atomic simulations. Actually, only the overall concentration $c(x)$ can be detected in experiments and atomic simulations. Moreover, it is likely to view the maximum concentration within the diffuse interface as the measured liquid concentration, despite it is inconsistent with that of determining the total driving force for diffuse interface, i.e., Eqs. (81)-(83). Therefore, we are currently defining $k(V)$ as

$$k(V) = \frac{\text{far-field concentration}}{c^{\max}} = \frac{c^{-\delta}}{c^{\max}}, \qquad (99)$$

which is also consistent with the Haapalehto et al.'s MD simulations [16].

Figure. 6(a) and (b) exhibit the composition profiles of Al-1 at. % Cu and Al-2 at.% Cu at different interface velocities, respectively. Their corresponding partition coefficients are displayed in Figure. 7, in which there is a satisfactory agreement with the MD results. In particular, with the considerations of long-range diffusion with relaxation effect by Eqs. (96)-(98), the solute profiles become homogeneous when interface velocity exceed the characteristic diffusion velocity [$V_c$], which is consistent with the MD simulated complete trapping at the finite velocity. In addition, it also proves the validity of the approximation of relaxation effect in Eq. (97), which may be adopted for future 2-D or 3-D studies.

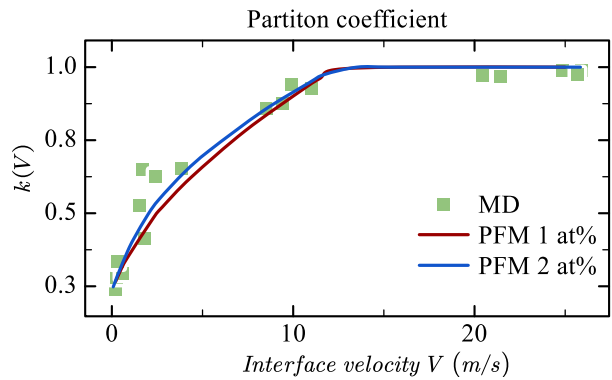

FIG. 7. Velocity-dependent partition coefficient predicted by Eqs. (33)-(37) and (96)-(98), in comparison with Haapalehto et al.'s MD simulations [16].

Compared with solute trapping, the solute drag concept is directly related to the thermodynamics of non-equilibrium interfaces, given by Eq. (84). The variations of $Q_m$ and $Q_d$ of Eq. (85) are exhibited in Figure. 8. It is shown that the evolving tendencies of $Q_m$ and $Q_d$ agree well with the interface friction and solute drag of the classical diffuse interface theories [56,57]: $Q_m$ linearly increases with $V_n$ while $Q_d$ increases first and then decreases. After reaching the characteristic diffusion velocity $V_c$, solute drag $Q_d = 0$ for complete solute trapping, which can be understood by the second expression in Eq. (84) and the homogeneous solute profile "$V \geq 12.5 \text{m/s}$" in Figure. 6. Moreover, when complete trapping occurs, the thermodynamic driving force is solely consumed by interface migration, i.e., diffusionless solidification. This is supported by the undercooling in Figure. 9. Note that the present simulations investigate the transient solidification kinetics, thus, interface velocity monotonically increases with the undercooling. The good agreement between transient kinetics (instead of steady kinetics) has also been found in Si-As system [12,13,81].

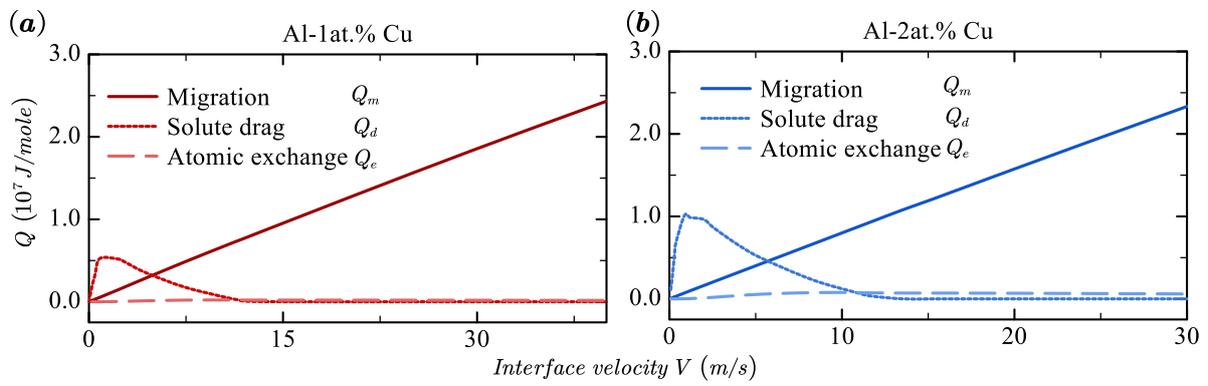

FIG. 8. Free energy dissipations of interface friction $Q_m$, solute drag $Q_d$ and part of atomic exchange $Q_e$.

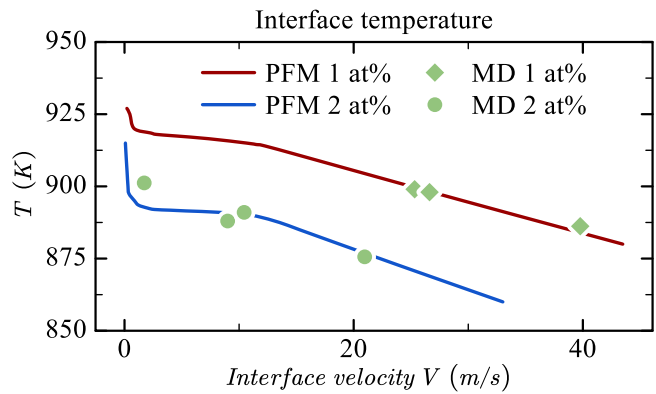

FIG. 9. Predicted interface temperature versus interface velocity.

It should note that $Q_e$ increases much slower with $V_n$, and its contribution to total dissipation can be ignored if in the high-velocity regime. However, it does not indicate than energy dissipation by (short-range) trans-sharp-interface diffusion is insignificant, because some part of its contribution has been incorporated in $Q_m$. Since $Q_e$ is proportionate to the difference in diffusion potentials, its negligible contribution indicates the jump of chemical potential at each point within the diffuse interface is relatively small throughout the whole velocity regime. In other words, even the far-from-equilibrium diffuse interface at the high-velocity regime is still composed of sharp interfaces (of RVEs) near the equilibrium. Since each point is comparable to the classical sharp interface models [24,25], we can define the velocity-dependent partition coefficient for each sharp interface of RVEs by the separated $c_s$ and $c_l$, i.e.,

$$k^*(V) = \frac{c_s(x)}{c_l(x)}. \tag{100}$$

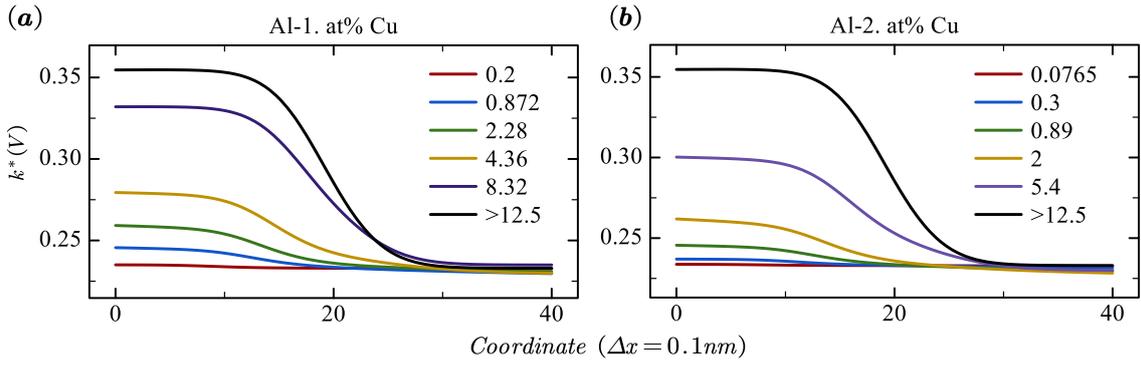

FIG. 10. Predicted velocity-dependent partition coefficient $k^*(V)$ [Eq. (100)] for each point inside the diffuse interface.

Figure. 10 displays the variations of $k^*(V)$ inside the diffuse interface at different interface velocities. Due to the spatial variation of $M^*$ in Eq. (32), trans-sharp-interface diffusion near the liquid boundary is significantly fast and its partition coefficient is close to the equilibrium $k_e$. In contrast, $k^*(V)$ at the solid boundary changes more significantly as an increase in velocities. At the low-velocity regime, $k^*(V)$ is close to $k_e$ through the whole diffuse interface, consistent with the basic assumption of the Kim-Kim-Suzuki model [53]. However, even at the high-velocity regime

33 / 52

where complete trapping occurs in Figure. 6, the $k^*(V)$ near the solid boundary is still not frome equilibrium $k_e$. It indicates that KKS model [53] is a good approximation of diffuse interface thermodynamics, and some properties of KKS may be useful for future quantitative extension of present model to a large size scale.

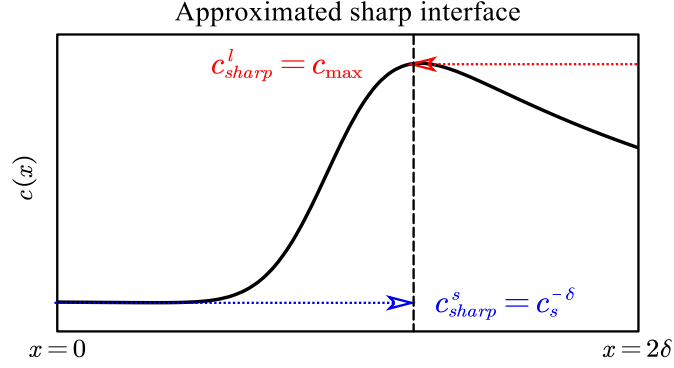

FIG. 11. Schematic diagram of an approximated sharp interface for the current diffuse interface.

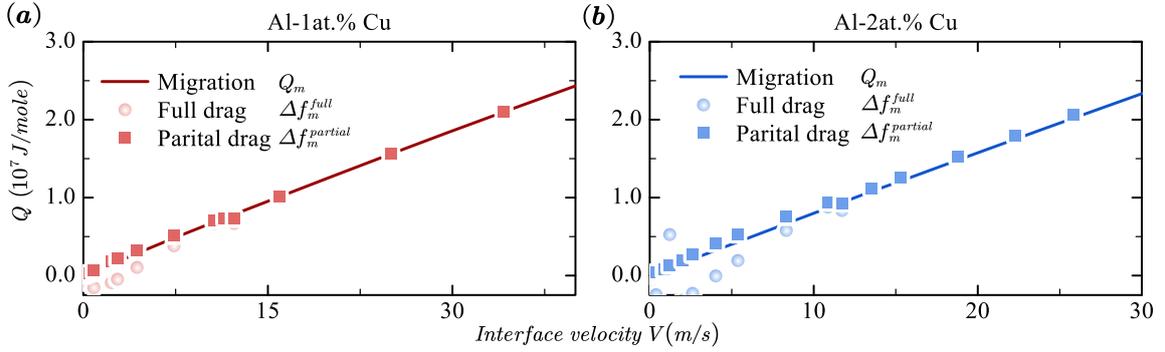

FIG. 12. Comparison of *full-drag* and *partial drag* models with the present dissipation by interface migration.

Finally, we can revisit the "partial drag" phenomenon in atomic simulations [14–18] with current phase-field model. As discussed before, it is likely to view the maximum concentration within the diffuse interface as the measured liquid concentration. Alternatively, there is an approximated sharp interface for present diffuse interface, as shown in Figure. 11, where $c_s^{sharp} = c^{-\delta}$ and $c_l^{sharp} = c^{max}$. As discussed in Eqs. (A11)-(A20) of the Appendix, although the classical sharp interface models based on the maximal entropy production principle [] can also be reinterpreted by the "two-part" mechanism revealed in Sec. III. B, its proportion of two parts is determined by the diffusion fluxes at the bulk boundaries, thus, it still yields the full-drag model for predicting driving force of interface migration (or friction), i.e.,

$$\Delta f_m^{full} = \frac{1}{v_m}\left[ f_l^{c_l^{sharp}} - f_s^{c_s^{sharp}} - \tilde{\mu}_s^{c_s^{sharp}}\left(c_l^{sharp} - c_s^{sharp}\right)\right], \tag{101}$$

whose deviation with $Q_m$ is shown in Figure. 12. In contrast, the driving force of interface migration in the *partial-drag* model [38] is usually given by

$$\Delta f_m^{partial} == \frac{1}{v_m}\left[ f_l^{c_l^{sharp}} - f_s^{c_s^{sharp}} - \frac{\tilde{\mu}_s^{c_s^{sharp}} + \tilde{\mu}_l^{c_l^{sharp}}}{2}\left(c_l^{sharp} - c_s^{sharp}\right)\right], \tag{102}$$

in which the phenomenological coefficient "1/2" was recently summarized by Antillon et al. [14] based on existing atomic simulations [14–18]. As compared in Figure. 12, Eq. (102) shows good consistency with the present model, which means the present model can reproduce the partial solute drag self-consistently in the classical irreversible thermodynamics, i.e., maximal entropy production principle [29–33]. More rigorously mathematic investigation of the consistency between the present phase-field model and Eq. (102), i.e., approximated sharp interface in Figure. 11, may be important for quantitative extension of present model into a more realistic scale for mesoscale microstructural pattern formation.

## V.  Conclusions

Assuming interface as a mixture of two phases with two conserved and two non-conserved concentration fields, we present a new phase-field model for non-equilibrium interface conditions within the framework of maximal entropy production principle. The critical novel feature is transparent dissipative processes inside the diffuse interface and the clarified balance of driving forces and energy dissipations, which bridges the classical sharp and diffuse interface theories with the phase-field method. As a practical application, the model is shown to represent the main characteristics of non-equilibrium interface kinetics, i.e., solute trapping and partial drag phenomena of rapid solidification. The main findings and conclusions are as follows:

(a) The present diffuse interface is an integral of numerous representative volume elements (RVEs) related by the (long-range) trans-diffuse interface diffusion. Each RVE has a sharp-interface phase transformation governed by phase-field migration and (short-range) trans-sharp-interface diffusion, comparable to classical sharp interface models. The proposed "two-part" mechanism explains the energy dissipations of displacive phase boundary migration and diffusional solute redistribution and clarifies the previous controversy about the classical sharp interface models. The similarity of the present phase transformation mechanism inside the interface with the recent experimental findings by neutron total scattering should also be noted.

(b) The present model is analytically traceable for the steady-state analysis. The classical diffuse interface theory's interface friction and solute drag are recovered by integrating the dissipations of three dissipative processes. Therefore, two main methods for non-equilibrium interface kinetics, i.e., the classical sharp and diffuse interface theories, are unified into the present phase-field model.

(c) With an approximated effective mobility for long-range diffusion, this work reproduces the transition from diffusional solidification to diffusionless solidification at finite interface velocity, after which there are complete solute trapping and disappearance of solute drag. The simulated free energy dissipation and non-equilibrium partition coefficient inside the diffuse interface indicate that even the far-from-equilibrium interface is still composed of the near-equilibrium sharp interfaces of RVEs. In addition, in the low-velocity regime with significant solute drag effects, the model self-consistently reproduces the MD-simulated partial drag phenomena within the maximal entropy production principle.

# Appendix: reinterpreted sharp interface model

This appendix reinterprets the previous sharp interface model [24–27], similarly used in the diffuse interface model of the main text. Figure A1 exhibits the 1-D planar sharp interface of a binary $\alpha-\gamma$ system at the constant temperature and volume, where two phases have the same molar volume for simplicity. The whole system is closed with no diffusion flux across the $\alpha$ or $\gamma$ surfaces ($J_{i=\alpha,\gamma}^{surf}=0$), while the separate phase is open at the interface with short-range diffusion fluxes of redistributing solute concentrations ($J_{i=\alpha,\gamma}^{*}\neq 0$) after interface migration ($V_n\neq 0$). To be consistent with the main text, $J_{i=\alpha,\gamma}^{*}$ is also referred to as the trans-sharp-interface diffusion in the following discussion.

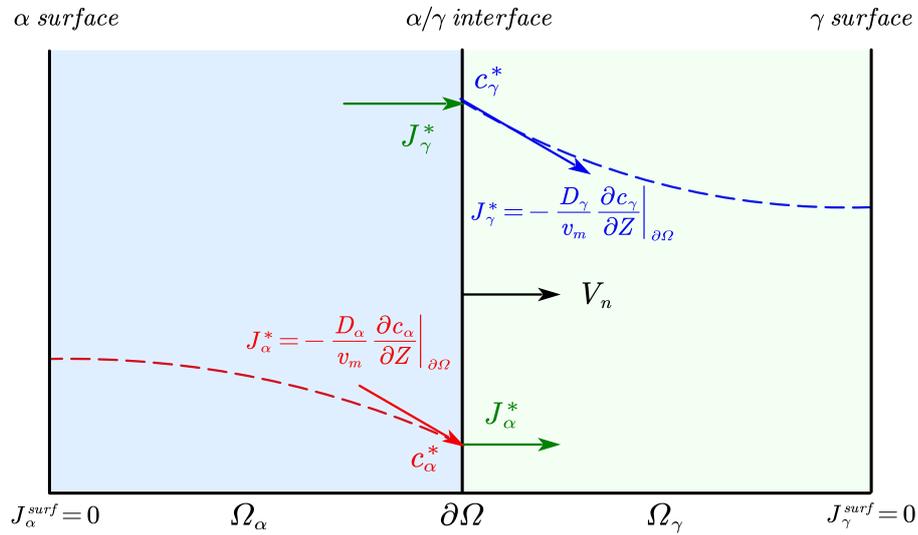

FIG. A1. Schematic diagram of the 1-D planar $\alpha-\gamma$ system. The left $\alpha$ phase and right $\gamma$ phase are separated by the sharp interface $\partial\Omega$ with a velocity $V_n$. The left and right boundaries are the surfaces of $\alpha$ and $\gamma$, respectively, where the surface fluxes $J_{i=\alpha,\gamma}^{surf}=0$ for a closed system. The red and blue dashed lines represent the composition profiles of $\alpha$ and $\gamma$ bulks, respectively. Approximated by the Fick's first law, the red and blue arrows denote the diffusion fluxes at the phase boundaries, which equal to those of trans-sharp-interface diffusion given by the green arrows.

In the sharp interface theory, the contributions of bulks and interface are separated, and the change rate of free energy at the interface ($\partial\Omega$) is generally given by

$$\dot{F}_{\partial\Omega} = \int_{\partial\Omega}\left[\frac{V_n}{v_m}\left(f_\alpha^* - f_\gamma^*\right) + J_\gamma^*\tilde{\mu}_\gamma^* - J_\alpha^*\tilde{\mu}_\alpha^*\right]d\partial\Omega, \tag{A1}$$

where $f^*_{i=\alpha,\gamma}$ is the free energy and $\tilde{\mu}^*_{i=\alpha,\gamma}$ is the diffusion potential. In Wang and Kuang et al.'s works [24–27], the evolution path of the system follows the maximum entropy production principle (MEPP) [29–33], i.e., $\dot{F}_{\partial\Omega} + Q_{\partial\Omega} = 0$. In the case of linear thermodynamics [32,33], the total energy dissipation at the interface $Q_{\partial\Omega}$ is usually expressed as a quadratic form of fluxes, i.e.,

$$Q_{\partial\Omega} = \int_{\partial\Omega} \left[ \frac{V_n^2}{M_V} + \frac{(J_\gamma^*)^2}{M_{J_\gamma^*}} + \frac{(J_\alpha^*)^2}{M_{J_\alpha^*}} \right] d\partial\Omega, \qquad (A2)$$

where $M_V$ and $M_{J_i^*}$ $(i=\alpha,\gamma)$ are the mobility for interface migration and trans-sharp-interface diffusion. In solid-state phase transformation, $M_V$ is usually assumed to follow the Arrhenius relation [26,28]. In contrast, $M_V = (V_0 v_m)/(RT)$ is normally adopted in solidification [24,25], where $V_0$ is the maximum possible transformation velocity. Based on the Aziz's chemical reaction rate theory of trans-interface diffusion [34,36], $M_{J_i^*}$ is generally defined by

$$M_{J_i^*} = \frac{D_i}{a} \left( \frac{\partial \tilde{\mu}_i}{\partial c_i} \right)^{-1} \quad i = \alpha, \gamma \qquad (A3)$$

where $D_i$ is the bulk diffusion coefficient and $a$ is the interatomic spacing [26,28].

In addition to the constraint of maximum entropy production, the mass conservation requires a Stefan's condition at the interface

$$. J_\alpha^* - J_\gamma^* = \frac{V_n}{v_m} (c_\alpha^* - c_\gamma^*). \qquad (A4)$$

where $c^*_{i=\alpha,\gamma}$ is the concentration at the interface (or called the concentration at phase boundary). Then, three fluxes according to MEPP follow [29–33] ($\lambda_{sharp}$ is the Lagrange multiplier for the constraint of Eq. (A4))

$$\delta \left\{ \dot{F}_{\partial\Omega} + \frac{Q_{\partial\Omega}}{2} + \lambda_{sharp} \left[ J_\alpha^* - J_\gamma^* - \frac{V_n}{v_m} (c_\alpha^* - c_\gamma^*) \right] \right\}_{V_n, J_\alpha^*, J_\gamma^*} = 0, \qquad (A5)$$

which yield as

$$\frac{V_n}{M_V} + \frac{\left(c_\alpha^* - c_\gamma^*\right)J_\alpha^*}{v_m M_{J_\alpha^*}} = \frac{1}{v_m}\left[f_\gamma^* - f_\alpha^* - \tilde{\mu}_\alpha^*\left(c_\gamma^* - c_\alpha^*\right)\right], \quad (A6)$$

$$J_\alpha^* = \frac{M_{J_\gamma^*} M_{J_\alpha^*}}{M_{J_\gamma^*} + M_{J_\alpha^*}}\left(\tilde{\mu}_\alpha^* - \tilde{\mu}_\gamma^*\right) + \frac{M_{J_\alpha^*}}{M_{J_\gamma^*} + M_{J_\alpha^*}}\frac{V_n}{v_m}\left(c_\alpha^* - c_\gamma^*\right), \quad (A7)$$

$$J_\gamma^* = \frac{M_{J_\gamma^*} M_{J_\alpha^*}}{M_{J_\gamma^*} + M_{J_\alpha^*}}\left(\tilde{\mu}_\alpha^* - \tilde{\mu}_\gamma^*\right) - \frac{M_{J_\gamma^*}}{M_{J_\gamma^*} + M_{J_\alpha^*}}\frac{V_n}{v_m}\left(c_\alpha^* - c_\gamma^*\right). \quad (A8)$$

The diffusion fluxes at the interface, $J_\alpha^*$ and $J_\gamma^*$, can be approximated by the Fick's first law,

$$J_\alpha^* = -\frac{D_\alpha}{v_m}\frac{\partial c_\alpha}{\partial Z}\bigg|_{Z=\partial\Omega}, \quad (A9)$$

$$J_\gamma^* = -\frac{D_\gamma}{v_m}\frac{\partial c_\gamma}{\partial Z}\bigg|_{Z=\partial\Omega}, \quad (A10)$$

in which $\partial c_i/\partial Z|_{Z=\partial\Omega}$ is the concentration gradient at the phase boundaries. Combining Eqs. (A6)-(A10) can determine $V_n$, $J_\alpha^*$, $J_\gamma^*$, $c_\alpha^*$, $c_\gamma^*$ at the same time, without priorly assuming the transformation direction.

Now, we can revisit the balance of driving forces and energy dissipation for different dissipative processes. Eq. (A4) can be reorganized as

$$J_{i=\alpha,\gamma}^* = \frac{V_n}{v_m}\left(c_i^* - c^{trans}\right), \quad (A11)$$

in which $c^{trans} = c_\alpha^* - v_m J_\alpha^*/V_n = c_\gamma^* - v_m J_\gamma^*/V_n$ is the composition of materials actually transferring the interface, as defined by Hillert et al. [21]. Submitting Eq. (A11) into Eq. (A1) and dividing $V_n/v_m$, the total (absolute) driving force of the sharp interface yields as

$$\Delta f_{tot}^{sharp} = f_\gamma^* - f_\alpha^* - \tilde{\mu}_\alpha^*\left(c^{trans} - c_\alpha\right) + \tilde{\mu}_\gamma^*\left(c^{trans} - c_\gamma\right). \quad (A12)$$

Firstly, let us consider a general case with nonnegligible diffusion in the growing phase, $J_\alpha^* \neq 0$, Eq. (A6) can be rewritten as

$$V_n = \frac{M_V^{eff}}{v_m}\left[f_\gamma^* - f_\alpha^* - \left(\frac{\tilde{\mu}_\alpha^* M_{J_\alpha^*} + \tilde{\mu}_\gamma^* M_{J_\gamma^*}}{M_{J_\alpha^*} + M_{J_\gamma^*}}\right)(c_\gamma^* - c_\alpha^*)\right];$$

$$\frac{1}{M_V^{eff}} = \frac{1}{M_V} + \frac{(c_\alpha^* - c_\gamma^*)^2}{v_m^2\left(M_{J_\alpha^*} + M_{J_\gamma^*}\right)},$$

(A13)

where $M_V^{eff}$ is an effective interface mobility similar to that of $\partial\phi/\partial t$ in the main text. From the averaged diffusion potential in Eq. (A13), one can see that the interface migration of the previous sharp-interface model also follows a "two-part" mechanism, as that of the phase field model in the main text. Then, the energy dissipation by interface migration $Q_m^{sharp}$ and trans-interface diffusion $Q_d^{sharp}$ can be reinterpreted currently as

$$Q_m^{sharp}\Big|_{J_\alpha^* \neq 0} = f_\gamma^* - f_\alpha^* - \left(\frac{\tilde{\mu}_\alpha^* M_{J_\alpha^*} + \tilde{\mu}_\gamma^* M_{J_\gamma^*}}{M_{J_\alpha^*} + M_{J_\gamma^*}}\right)(c_\gamma^* - c_\alpha^*),$$

(A14)

$$Q_d^{sharp}\Big|_{J_\alpha^* \neq 0} = \left(\tilde{\mu}_\alpha^* - \tilde{\mu}_\gamma^*\right)\left(\frac{M_{J_\alpha^*} c_\gamma^* + M_{J_\gamma^*} c_\alpha^*}{M_{J_\alpha^*} + M_{J_\gamma^*}} - c^{trans}\right),$$

(A15)

which are different from previous works by Hillert et al. [19–21] and Wang and Kuang et al. [26].

Then, we consider a special case with negligible diffusion in growing phase, $J_\alpha^* = 0, D_\gamma \gg D_\alpha = 0$, which is generally valid for solidification ($\alpha = solid$, $\gamma = liquid$). Taking $J_\alpha^* = 0$ into Eqs. (A6)-(A8) and (A12), there will be

$$\Delta f_{tot}^{sharp}\Big|_{J_\alpha^* = 0} = f_\gamma^* - f_\alpha^* - \tilde{\mu}_\gamma^*(c_\gamma^* - c_\alpha^*),$$

(A16)

$$V_n\Big|_{J_\alpha^* = 0} = \frac{M_V}{v_m}\left[f_\gamma^* - f_\alpha^* - \tilde{\mu}_\alpha^*(c_\gamma^* - c_\alpha^*)\right],$$

(A17)

$$J_\gamma^*\Big|_{J_\alpha^* = 0} = M_{J_\gamma^*}\left(\tilde{\mu}_\alpha^* - \tilde{\mu}_\gamma^*\right) = -\frac{V_n}{v_m}(c_\alpha^* - c_\gamma^*),$$

(A18)

from which one can obtain the energy dissipation in the classical *full-drag* model [19–28]

$$Q_m^{sharp}\Big|_{J_\alpha^* = 0} = f_\gamma^* - f_\alpha^* - \tilde{\mu}_\alpha^*(c_\gamma^* - c_\alpha^*),$$

(A19)

$$Q_d^{sharp}\big|_{J_\alpha^*=0} = \left(\tilde{\mu}_\alpha^* - \tilde{\mu}_\gamma^*\right)\left(c_\gamma^* - c_\alpha^*\right). \tag{A20}$$

It means that the *full-drag* is natural product of the MEPP [29–33], despite its predictions significantly deviate from the atomic simulations [14–18].

## Acknowledgement

The work was supported by the Research Fund of the State Key Laboratory of Solidification Processing (NPU), China (Grant No. 2020-TS-06, 2021-TS-02) and Natural Science Basic Research of Shannxi (Program No. 2022JC-28). Y. L. sincerely thanks Prof. Haifeng Wang in Northwestern Polytechnical University for the detailed introduction of the rapid solidification kinetics and his sharp and diffuse interface models.

## References

[1] D. Herzog, V. Seyda, E. Wycisk, and C. Emmelmann, *Additive Manufacturing of Metals*, Acta Mater. **117**, 371 (2016).

[2] W. J. Sames, F. A. List, S. Pannala, R. R. Dehoff, and S. S. Babu, *The Metallurgy and Processing Science of Metal Additive Manufacturing*, Int. Mater. Rev. **61**, 315 (2016).

[3] T. DebRoy, H. L. Wei, J. S. Zuback, T. Mukherjee, J. W. Elmer, J. O. Milewski, A. M. Beese, A. Wilson-Heid, A. De, and W. Zhang, *Additive Manufacturing of Metallic Components – Process, Structure and Properties*, Prog. Mater. Sci. **92**, 112 (2018).

[4] W. Wolf, L. P. Luiz, G. Zepon, C. S. Kiminami, C. Bolfarini, and W. J. Botta, *Single Step Fabrication by Spray Forming of Large Volume Al-Based Composites Reinforced with Quasicrystals*, Scr. Mater. **181**, 86 (2020).

[5] J. P. Oliveira, T. G. Santos, and R. M. Miranda, *Revisiting Fundamental Welding Concepts to*

*Improve Additive Manufacturing: From Theory to Practice*, Prog. Mater. Sci. **107**, 100590 (2020).

[6]  D. M. Herlach, *Metastable Solids from Undercooled Melts*, Mater. Sci. Forum **539–543**, 1977 (2007).

[7]  S. C. Gillf and W. Kurz, *Pergamon RAPIDLY SOLIDIFIED Al--Cu ALLOYS--II . C A L C U L A T I O N OF THE M I C R O S T R U C T U R E SELECTION MAP*, **43**, 139 (1995).

[8]  M. Gremaud, M. Carrard, and W. Kurz, *Banding Phenomena in AlFe Alloys Subjected to Laser Surface Treatment*, Acta Metall. Mater. **39**, 1431 (1991).

[9]  S. C. Gill and W. Kurz, *Rapidly Solidified AlCu Alloys-I. Experimental Determination of the Microstructure Selection Map*, Acta Metall. Mater. **41**, 3563 (1993).

[10] J. T. McKeown, A. K. Kulovits, C. Liu, K. Zweiacker, B. W. Reed, T. Lagrange, J. M. K. Wiezorek, and G. H. Campbell, *In Situ Transmission Electron Microscopy of Crystal Growth-Mode Transitions during Rapid Solidification of a Hypoeutectic Al-Cu Alloy*, Acta Mater. **65**, 56 (2014).

[11] J. A. Kittl, R. Reitano, M. J. Aziz, D. P. Brunco, and M. O. Thompson, *Time-Resolved Temperature Measurements during Rapid Solidification of Si-As Alloys Induced by Pulsed-Laser Melting*, J. Appl. Phys. **73**, 3725 (1993).

[12] J. A. Kittl, M. J. Aziz, D. P. Brunco, and M. O. Thompson, *Nonequilibrium Partitioning during Rapid Solidification of Si As Alloys*, J. Cryst. Growth **148**, 172 (1995).

[13] J. A. Kittl, P. G. Sanders, M. J. Aziz, D. P. Brunco, and M. O. Thompson, *Complete Experimental Test of Kinetic Models for Rapid Alloy Solidification*, Acta Mater. **48**, 4797 (2000).

[14]   E. A. Antillon, C. A. Hareland, and P. W. Voorhees, *Solute Trapping and Solute Drag during Non-Equilibrium Solidification of Fe–Cr Alloys*, Acta Mater. **248**, 118769 (2023).

[15]   H. Humadi, J. J. Hoyt, and N. Provatas, *Microscopic Treatment of Solute Trapping and Drag*, Phys. Rev. E **93**, 1 (2016).

[16]   M. Haapalehto, T. Pinomaa, L. Wang, and A. Laukkanen, *An Atomistic Simulation Study of Rapid Solidification Kinetics and Crystal Defects in Dilute Al–Cu Alloys*, Comput. Mater. Sci. **209**, 111356 (2022).

[17]   Y. Yang, H. Humadi, D. Buta, B. B. Laird, D. Sun, J. J. Hoyt, and M. Asta, *Atomistic Simulations of Nonequilibrium Crystal-Growth Kinetics from Alloy Melts*, Phys. Rev. Lett. **107**, 8 (2011).

[18]   S. Kavousi, B. R. Novak, J. Hoyt, and D. Moldovan, *Interface Kinetics of Rapid Solidification of Binary Alloys by Atomistic Simulations: Application to Ti-Ni Alloys*, Comput. Mater. Sci. **184**, (2020).

[19]   M. Hillert, *Solute Drag, Solute Trapping and Diffusional Dissipation of Gibbs Energy*, Acta Mater. **47**, 4481 (1999).

[20]   M. Hillert and M. Rettenmayr, *Deviation from Local Equilibrium at Migrating Phase Interfaces*, Acta Mater. **51**, 2803 (2003).

[21]   M. Hillert, J. Odqvist, and J. Ågren, *Interface Conditions during Diffusion-Controlled Phase Transformations*, Scr. Mater. **50**, 547 (2004).

[22]   M. Buchmann and M. Rettenmayr, *Rapid Solidification Theory Revisited - A Consistent Model Based on a Sharp Interface*, Scr. Mater. **57**, 169 (2007).

[23]   M. Buchmann and M. Rettenmayr, *Non-Equilibrium Transients during Solidification - A*

*Numerical Study*, Scr. Mater. **58**, 106 (2008).

[24] H. Wang, F. Liu, H. Zhai, and K. Wang, *Application of the Maximal Entropy Production Principle to Rapid Solidification: A Sharp Interface Model*, Acta Mater. **60**, 1444 (2012).

[25] K. Wang, H. Wang, F. Liu, and H. Zhai, *Modeling Rapid Solidification of Multi-Component Concentrated Alloys*, Acta Mater. **61**, 1359 (2013).

[26] W. Kuang, H. Wang, J. Zhang, and F. Liu, *Application of the Thermodynamic Extremal Principle to Diffusion-Controlled Phase-Transformations in Multi-Component Substitutional Alloys: Modeling and Applications*, Acta Mater. **120**, 415 (2016).

[27] W. Kuang, H. Wang, X. Li, J. Zhang, Q. Zhou, and Y. Zhao, *Application of the Thermodynamic Extremal Principle to Diffusion-Controlled Phase Transformations in Fe-C-X Alloys: Modeling and Applications*, Acta Mater. **159**, 16 (2018).

[28] J. Zhang, H. Wang, W. Kuang, Y. Zhang, S. Li, Y. Zhao, and D. M. Herlach, *Rapid Solidification of Non-Stoichiometric Intermetallic Compounds: Modeling and Experimental Verification*, Acta Mater. **148**, 86 (2018).

[29] L. Onsager, *Reciprocal Relations in Irreversible Processes. I.*, Phys. Rev. **37**, 405 (1931).

[30] L. Onsager, *Reciprocal Relations in Irreversible Processes. II.*, Phys. Rev. **38**, 2265 (1931).

[31] H. Ziegler, *An Introduction to Thermomechanics* (Elsevier, 2012).

[32] J. Svoboda, I. Turek, and F. D. Fischer, *Application of the Thermodynamic Extremal Principle to Modeling of Thermodynamic Processes in Material Sciences*, Philos. Mag. **85**, 3699 (2005).

[33] F. D. Fischer, J. Svoboda, and H. Petryk, *Thermodynamic Extremal Principles for Irreversible Processes in Materials Science*, Acta Mater. **67**, 1 (2014).

[34] M. J. Aziz, *Model for Solute Redistribution during Rapid Solidification*, J. Appl. Phys. **53**,

1158 (1982).

[35] M. J. Aziz, *Dissipation-Theory Treatment of the Transition from Diffusion-Controlled to Diffusionless Solidification*, Appl. Phys. Lett. **43**, 552 (1983).

[36] M. J. Aziz and T. Kaplan, *Continuous Growth Model for Interface Motion during Alloy Solidification*, Acta Metall. **36**, 2335 (1988).

[37] M. J. Aziz and W. J. Boettinger, *On the Transition from Short-Range Diffusion-Limited to Collision-Limited Growth in Alloy Solidification*, Acta Metall. Mater. **42**, 527 (1994).

[38] C. A. Hareland, G. Guillemot, C. A. Gandin, and P. W. Voorhees, *The Thermodynamics of Non-Equilibrium Interfaces during Phase Transformations in Concentrated Multicomponent Alloys*, Acta Mater. **241**, 118407 (2022).

[39] N. Provatas and K. Elder, *Phase-Field Methods in Materials Science and Engineering* (John Wiley \& Sons, 2011).

[40] N. V. Menshutina, A. V. Kolnoochenko, and E. A. Lebedev, *Cellular Automata in Chemistry and Chemical Engineering*, Annu. Rev. Chem. Biomol. Eng. **11**, 87 (2020).

[41] F. Gibou, R. Fedkiw, and S. Osher, *A Review of Level-Set Methods and Some Recent Applications*, J. Comput. Phys. **353**, 82 (2018).

[42] C. A. Becker, M. Asta, J. J. Hoyt, and S. M. Foiles, *Equilibrium Adsorption at Crystal-Melt Interfaces in Lennard-Jones Alloys*, J. Chem. Phys. **124**, 164708 (2006).

[43] D. Buta, M. Asta, and J. J. Hoyt, *Atomistic Simulation Study of the Structure and Dynamics of a Faceted Crystal-Melt Interface*, Phys. Rev. E **78**, 31605 (2008).

[44] C. A. Becker, D. L. Olmsted, M. Asta, J. J. Hoyt, and S. M. Foiles, *Atomistic Simulations of Crystal-Melt Interfaces in a Model Binary Alloy: Interfacial Free Energies, Adsorption

*Coefficients, and Excess Entropy*, Phys. Rev. B **79**, 54109 (2009).

[45] T. Frolov and Y. Mishin, *Solid-Liquid Interface Free Energy in Binary Systems: Theory and Atomistic Calculations for the (110) Cu--Ag Interface*, J. Chem. Phys. **131**, 54702 (2009).

[46] L.-Q. Chen, *Phase-Field Models for Microstructure Evolution*, Annu. Rev. Mater. Res. **32**, 113 (2002).

[47] N. Moelans, B. Blanpain, and P. Wollants, *An Introduction to Phase-Field Modeling of Microstructure Evolution*, Calphad **32**, 268 (2008).

[48] I. Steinbach, *Phase-Field Models in Materials Science*, Model. Simul. Mater. Sci. Eng. **17**, 73001 (2009).

[49] I. Steinbach, *Phase-Field Model for Microstructure Evolution at the Mesoscopic Scale*, Annu. Rev. Mater. Res. **43**, 89 (2013).

[50] L.-Q. Chen and Y. Zhao, *From Classical Thermodynamics to Phase-Field Method*, Prog. Mater. Sci. **124**, 100868 (2022).

[51] D. Tourret, H. Liu, and J. LLorca, *Phase-Field Modeling of Microstructure Evolution: Recent Applications, Perspectives and Challenges*, Prog. Mater. Sci. **123**, 100810 (2022).

[52] J. Tiaden, B. Nestler, H.-J. Diepers, and I. Steinbach, *The Multiphase-Field Model with an Integrated Concept for Modelling Solute Diffusion*, Phys. D Nonlinear Phenom. **115**, 73 (1998).

[53] S. G. Kim, W. T. Kim, and T. Suzuki, *Phase-Field Model for Binary Alloys*, Phys. Rev. E **60**, 7186 (1999).

[54] I. Steinbach, L. Zhang, and M. Plapp, *Phase-Field Model with Finite Interface Dissipation*, Acta Mater. **60**, 2689 (2012).

[55] L. Zhang and I. Steinbach, *Phase-Field Model with Finite Interface Dissipation: Extension to Multi-Component Multi-Phase Alloys*, Acta Mater. **60**, 2702 (2012).

[56] J. W. Cahn, *The Impurity-Drag Effect in Grain Boundary Motion*, Acta Metall. **10**, 789 (1962).

[57] M. Hillert and B. O. Sundman, *A Treatment of the Solute Drag on Moving Grain Boundaries and Phase Interfaces in Binary Alloys*, Acta Metall. **24**, 731 (1976).

[58] H. Wang, F. Liu, G. J. Ehlen, and D. M. Herlach, *Application of the Maximal Entropy Production Principle to Rapid Solidification: A Multi-Phase-Field Model*, Acta Mater. **61**, 2617 (2013).

[59] H. Wang, X. Zhang, C. Lai, W. Kuang, and F. Liu, *Thermodynamic Principles for Phase-Field Modeling of Alloy Solidification*, Curr. Opin. Chem. Eng. **7**, 6 (2015).

[60] O. Pierre-Louis, *Phase Field Models for Step Flow*, Phys. Rev. E **68**, 21604 (2003).

[61] Y. Cui, Y. Zhang, L. Sun, M. Feygenson, M. Fan, X. L. Wang, P. K. Liaw, I. Baker, and Z. Zhang, *Phase Transformation via Atomic-Scale Periodic Interfacial Energy*, Mater. Today Phys. **24**, 100668 (2022).

[62] F. D. Fischer and J. Svoboda, *Diffusion of Elements and Vacancies in Multi-Component Systems*, Prog. Mater. Sci. **60**, 338 (2014).

[63] J. Svoboda, F. D. Fischer, P. Fratzl, and E. Kozeschnik, *Modelling of Kinetics in Multi-Component Multi-Phase Systems with Spherical Precipitates: I: Theory*, Mater. Sci. Eng. A **385**, 166 (2004).

[64] F. D. Fischer, K. Hackl, and J. Svoboda, *Improved Thermodynamic Treatment of Vacancy-Mediated Diffusion and Creep*, Acta Mater. **108**, 347 (2016).

[65] Y. Li, Z. Wang, Y. Wang, J. Li, and J. Wang, *Revealing Curvature and Stochastic Effects on Grain Growth: A Thermodynamic Perspective from Extremal Principle*, Scr. Mater. **217**, 114766 (2022).

[66] Y. Li, Z. Wang, X. Gao, Y. Wang, J. Li, and J. Wang, *Revisiting Transient Coarsening Kinetics: A New Framework in the Lifshitz-Slyozov-Wagner Space*, Acta Mater. **237**, 118196 (2022).

[67] S. M. Allen and J. W. Cahn, *A Microscopic Theory for Antiphase Boundary Motion and Its Application to Antiphase Domain Coarsening*, Acta Metall. **27**, 1085 (1979).

[68] J. W. Cahn, *On Spinodal Decomposition*, Acta Metall. **9**, 795 (1961).

[69] J. Svoboda, F. D. Fischer, and D. L. McDowell, *Derivation of the Phase Field Equations from the Thermodynamic Extremal Principle*, Acta Mater. **60**, 396 (2012).

[70] A. Fang and Y. Mi, *Recovering Thermodynamic Consistency of the Antitrapping Model: A Variational Phase-Field Formulation for Alloy Solidification*, Phys. Rev. E **87**, 12402 (2013).

[71] A. A. Wheeler, W. J. Boettinger, and G. B. McFadden, *Phase-Field Model for Isothermal Phase Transitions in Binary Alloys*, Phys. Rev. A **45**, 7424 (1992).

[72] A. Mukherjee, J. A. Warren, and P. W. Voorhees, *A Quantitative Variational Phase Field Framework*, ArXiv Prepr. ArXiv2303.09671 (2023).

[73] Y. Li, L. Wang, J. Li, J. Wang, and Z. Wang, *Thermodynamics of Non-Equilibrium Diffuse-Interfaces in Mesoscale Phase Transformations*, ArXiv Prepr. ArXiv2303.09879 (2023).

[74] K. Karayagiz et al., *Finite Interface Dissipation Phase Field Modeling of Ni--Nb under Additive Manufacturing Conditions*, Acta Mater. **185**, 320 (2020).

[75] M. Hillert, *An Application of Irreversible Thermodynamics to Diffusional Phase*

*Transformations*, Acta Mater. **54**, 99 (2006).

[76] S. K. D. Nath, Y. Shibuta, M. Ohno, T. Takaki, and T. Mohri, *A Molecular Dynamics Study of Partitionless Solidification and Melting of Al--Cu Alloys*, ISIJ Int. **57**, 1774 (2017).

[77] S. L. Sobolev, *Local-Nonequilibrium Model for Rapid Solidification of Undercooled Melts*, Phys. Lett. A **199**, 383 (1995).

[78] S. L. Sobolev, *Rapid Solidification under Local Nonequilibrium Conditions*, Phys. Rev. E **55**, 6845 (1997).

[79] S. L. Sobolev, *Local Non-Equilibrium Diffusion Model for Solute Trapping during Rapid Solidification*, Acta Mater. **60**, 2711 (2012).

[80] S. L. Sobolev, *On the Transition from Diffusion-Limited to Kinetic-Limited Regimes of Alloy Solidification*, Acta Mater. **61**, 7881 (2013).

[81] S. L. Sobolev, *A Novel Hybrid Model Combining Continuum Local Nonequilibrium and Discrete Variables Methods for Solute Trapping during Rapid Alloy Solidification*, Acta Mater. **116**, 212 (2016).

[82] P. Galenko and S. Sobolev, *Local Nonequilibrium Effect on Undercooling in Rapid Solidification of Alloys*, Phys. Rev. E **55**, 343 (1997).

[83] P. Galenko, *Extended Thermodynamical Analysis of a Motion of the Solid-Liquid Interface in a Rapidly Solidifying Alloy*, Phys. Rev. B **65**, 144103 (2002).

[84] D. M. Herlach and P. K. Galenko, *Rapid Solidification: In Situ Diagnostics and Theoretical Modelling*, Mater. Sci. Eng. A **449**–**451**, 34 (2007).

[85] P. K. Galenko and M. D. Krivilyov, *Modeling of a Transition to Diffusionless Dendritic Growth in Rapid Solidification of a Binary Alloy*, Comput. Mater. Sci. **45**, 972 (2009).

[86] P. K. Galenko, E. V Abramova, D. Jou, D. A. Danilov, V. G. Lebedev, and D. M. Herlach, *Solute Trapping in Rapid Solidification of a Binary Dilute System: A Phase-Field Study*, Phys. Rev. E **84**, 41143 (2011).

[87] H. Wang, P. K. Galenko, X. Zhang, W. Kuang, F. Liu, and D. M. Herlach, *Phase-Field Modeling of an Abrupt Disappearance of Solute Drag in Rapid Solidification*, Acta Mater. **90**, 282 (2015).

[88] A. A. Wheeler, W. J. Boettinger, and G. B. McFadden, *Phase-Field Model of Solute Trapping during Solidification*, Phys. Rev. E **47**, 1893 (1993).

[89] N. A. Ahmad, A. A. Wheeler, W. J. Boettinger, and G. B. McFadden, *Solute Trapping and Solute Drag in a Phase-Field Model of Rapid Solidification*, Phys. Rev. E **58**, 3436 (1998).

[90] S. Kavousi and M. A. Zaeem, *Quantitative Phase-Field Modeling of Solute Trapping in Rapid Solidification*, Acta Mater. **205**, 116562 (2021).

[91] D. Danilov and B. Nestler, *Phase-Field Modelling of Solute Trapping during Rapid Solidification of a Si--As Alloy*, Acta Mater. **54**, 4659 (2006).

[92] T. Pinomaa and N. Provatas, *Quantitative Phase Field Modeling of Solute Trapping and Continuous Growth Kinetics in Quasi-Rapid Solidification*, Acta Mater. **168**, 167 (2019).

[93] H. Wang, F. Liu, W. Yang, Z. Chen, G. Yang, and Y. Zhou, *Solute Trapping Model Incorporating Diffusive Interface*, Acta Mater. **56**, 746 (2008).

[94] V. G. Lebedev, E. V Abramova, D. A. Danilov, and P. K. Galenko, *Phase-Field Modeling of Solute Trapping: Comparative Analysis of Parabolic and Hyperbolic Models*, Int. J. Mater. Res. **101**, 473 (2010).

[95] L. Zhang, E. V. Danilova, I. Steinbach, D. Medvedev, and P. K. Galenko, *Diffuse-Interface*

*Modeling of Solute Trapping in Rapid Solidification: Predictions of the Hyperbolic Phase-Field Model and Parabolic Model with Finite Interface Dissipation*, Acta Mater. **61**, 4155 (2013).